\documentclass[aps,pra,reprint,twocolumn,superscriptaddress,showkeys,a4paper]{revtex4}
\usepackage{amsfonts}
\usepackage{amssymb}
\usepackage{bm, bbm, dsfont}
\usepackage[dvips]{graphicx}
\usepackage[T1]{fontenc}
\usepackage[latin2]{inputenc}
\usepackage{amsmath}
\usepackage{latexsym}

\usepackage{color}

\def\dt{{\rm d}\,}

\newcommand{\ket}[1]{| #1 \rangle}
\newcommand{\bra}[1]{\langle #1 |}

\def\duzomniejsze{<\kern-.7mm<}
\def\duzowieksze{>\kern-.7mm>}

\def\textbf#1{{\bf #1}}
\def\be{\begin{equation}}
\def\ee{\end{equation}}
\def\ben{\begin{eqnarray}}
\def\een{\end{eqnarray}}
 \def\beqa{\begin{eqnarray}}
\def\eeqa{\end{eqnarray}}
\def\eea{\end{array}}
\def\bea{\begin{array}}
\newcommand{\bei}{\begin{itemize}}
\newcommand{\eei}{\end{itemize}}
\newcommand{\bee}{\begin{enumerate}}
\newcommand{\eee}{\end{enumerate}}

\def\hcal{{\cal H}}

\def\1{\openone}

\def\tr{{\rm Tr}}

\def\>{\rangle}
\def\<{\langle}

\def\dt#1{{{\kern -.0mm\rm d}}#1\,}
\def\squareforqed{\hbox{\rlap{$\sqcap$}$\sqcup$}}
\def\qed{\ifmmode\squareforqed\else{\unskip\nobreak\hfil
\penalty50\hskip1em\null\nobreak\hfil\squareforqed
\parfillskip=0pt\finalhyphendemerits=0\endgraf}\fi}

\def\supp{{\rm supp\,}}

\newtheorem{lemma}{Lemma}
\newtheorem{theorem}[lemma]{Theorem}
\newtheorem{main result}[lemma]{Main result}
\newtheorem{proposition}[lemma]{Proposition}
\newtheorem{definition}{Definition}

\newtheorem{fact}[lemma]{Fact}

\def\bep{\begin{proposition}}
\def\eep{\end{proposition}}
\def\bel{\begin{lemma}}
\def\eel{\end{lemma}}

\def\bet{\begin{theorem}}
\def\eet{\end{theorem}}
\def\bed{\begin{definition}}
\def\eed{\end{definition}}
\def\bef{\begin{fact}}
\def\eef{\end{fact}}

\begin{document}

\title{Objectivity Through State Broadcasting: The Origins Of Quantum Darwinism}

\author{J.~K.~Korbicz}
 \email{jaroslaw.korbicz@ug.edu.pl}
  \affiliation{Institute of Theoretical Physics and Astrophysics, University of Gda\'nsk, 80-952 Gda\'nsk, Poland}
 \affiliation{National Quantum Information Centre in Gdan\'sk, 81-824 Sopot, Poland}
\author{P.~Horodecki}
\affiliation{Faculty of Applied Physics and Mathematics, Gda\'nsk University of Technology, 80-233 Gda\'nsk, Poland}
\affiliation{National Quantum Information Centre in Gdan\'sk, 81-824 Sopot, Poland}
\author{R.~Horodecki}
\affiliation{Institute of Theoretical Physics and Astrophysics, University of Gda\'nsk, 80-952 Gda\'nsk, Poland}
\affiliation{National Quantum Information Centre in Gdan\'sk, 81-824 Sopot, Poland}

\date{\today}

\begin{abstract}
Quantum mechanics is one of the most successful theories, correctly predicting huge class of physical phenomena. 
Ironically, in spite of all its successes, there is a notorious problem: how does Nature create a ''bridge'' from fragile quanta 
to the robust, objective world of everyday experience? 
It is now commonly accepted that the most promising approach is the Decoherence Theory, based on the system-environment paradigm. 
To explain the observed redundancy and 
objectivity of information in the classical realm, Zurek proposed to divide the environment into independent fractions 
and argued that each of them carries a nearly complete classical information about the system. This Quantum Darwinism model has nevertheless  some serious drawbacks:
i) the entropic information redundancy is motivated by a priori purely classical reasoning; ii) there is no answer to the basic question: 
what physical process makes the transition from quantum description to classical objectivity possible?  Here we prove that the 
necessary and sufficient condition for objective existence of a state is the spectrum broadcasting process, which, in particular, implies Quantum Darwinism. 
We first show it in general, using multiple environments
paradigm, a suitable definition of objectivity, and Bohr's  notion of non-disturbance,
and then on  the emblematic example for Decoherence Theory:  a dielectric sphere illuminated by photons. 
We also apply  Perron-Frobenius Theorem to show a faithful, ''decoherence-free'' form of broadcasting. 
We suggest that the spectrum broadcasting might be one of the foundational properties of 
Nature, which opens a ''window'' for life processes. 
\end{abstract}


\maketitle
\section{Introduction}

Uninterrupted series of successes of quantum mechanics support a belief that quantum formalism applies to all of physical reality. Thus, in particular, the objective classical world 
of everyday experience should emerge naturally from the formalism. This has been a long-standing problem,
in fact already present from the very dawn of quantum mechanics 
(see e.g. the writings of Bohr \cite{old_bohr} and Heisenberg \cite{old_heis} for some of the earlier discussions and e.g. \cite{modern} for some of the modern approaches, relevant to the present work).  
Perhaps the most promising approach is Decoherence Theory (see e.g. \cite{decoh}), based on a system-environment paradigm: a quantum system is considered not in an isolation, but rather interacting 
with its environment.  
It  recovers, under certain conditions, a classical-like behavior of the system alone in some preferred frame, singled out by the interaction and 
called a \emph{pointer basis}, and explains it through information leakage from the system to the environment (the system is ''monitored'' by its environment).

However, as Zurek  noticed recently \cite{ZurekNature}, Decoherence Theory is silent on how comes that in the classical 
realm information is \emph{redundant}---same record 
can exist in a large number of copies and can be independently accessed by many observers and many times.
To overcome the problem, he has introduced a more realistic model of environment, composed of a number of independent fractions, and argued using several models
(see e.g. Refs.~\cite{ZurekPRL,Zurekspins}) that after the decoherence has taken place, each of these fractions carries a nearly complete classical information about the system. 
Then Zurek argues that this huge information redundancy implies  objective existence \cite{ZurekNature}.
This model, called Quantum Darwinism, although very attractive (see Ref.~\cite{exp} for some experimental evidence), 
has a certain gap which make its foundations not very clear. 
Postponing the details to Section \ref{qdcond}, the criterion used in Quantum Darwinism 
to show the information redundancy is motivated by \emph{entirely classical}
reasoning and a priori may not work as intended in the quantum world.

There is however another basic question: is there a fundamental physical process, consistent with the laws of quantum mechanics, which leads to the appearance in the environment of multiple copies of a state of the system?  
In other words,  how does Nature create a ''bridge'' from fragile quantum states, which cannot be cloned \cite{no_cloning}, to robust classical objectivity? 
Zurek is aware of the difficulty when he writes \cite{ZurekNature}: 
\begin{quote}
\emph{''Quantum Darwinism leads to appearance in the environment of multiple copies of the state of the system. 
However the no-cloning theorem prohibits copying of unknown quantum states.''}
\end{quote} 
However, he does not provide a clear answer to the question \cite{ZurekNature}:
\begin{quote}
\emph{''Quick answer is that cloning refers to (unknown) quantum states. 
So, copying of observables evades the theorem. Nevertheless, the tension between the prohibition on cloning and the need for copying is revealing: It leads to breaking of unitary symmetry implied by the superposition principle, [...]''}
\end{quote} 

But the no-cloning theorem prohibits only \emph{uncorrelated} copies of the state of the system,  whereas  it leaves open a possibility of producing \emph{correlated} ones. 
This is the essence of \emph{state broadcasting}---a process aimed at proliferating a given state through correlated copies \cite{broadcasting}. 
In this work we identify a weaker form of state broadcasting---\emph{spectrum broadcasting}, introduced in Ref.~\cite{my},
as the fundamental physical process, consistent with quantum mechanical laws, 
which leads to the perceived objectivity of classical information, and as a result recover Quantum Darwinism (as a limiting point).  
We do it first in full generality, using a definition of objective existence due to Zurek \cite{ZurekNature} and Bohr's notion of 
non-disturbance \cite{Bohr, Wiseman}. Then, in one of the emblematic examples of Decoherence Theory and Quantum Darwinism: 
a small dielectric sphere illuminated by photons (see e.g. Refs.~\cite{ZurekPRL,JoosZeh,GallisFleming,HornbergerSipe,RiedelZurek}). 
The recognition of the underlying spectrum broadcasting mechanism has been possible due to a paradigmatic shift in the core object of the analysis. 
From a partial state of the system (Decoherence Theory) or information-theoretical quantities like mutual information (Quantum Darwinism)
to a full quantum state of the system and the observed environment. This also opens a possibility for direct experimental tests using e.g. 
quantum state tomography \cite{tomo}.

\section{Objective existence needs state broadcasting}
\label{tw}

What does it mean that something \emph{objectively exists}? What does it mean for information? 
For the purpose of this study we employ the definition from Ref.~\cite{ZurekNature}:
\begin{definition}[Objectivity]\label{obj}
A state of the system $S$ 
exists objectively if ''[...]many observers can find out the state of $S$ independently, and without perturbing it.'' 
\end{definition}

In what follows we will try to make this definition as precise as possible  and investigate its consequences.
The natural setting for this is Quantum Darwinism \cite{ZurekNature}: the quantum system of interest $S$ interacts with multiple 
environments $E_1,\dots,E_N$ (denoted collectively as $E$), also modeled as quantum systems. 
The environments (or their collections) are monitored by independent observers (\emph{environmental observers})
and here we do not assume symmetric environments---they can be all different. 
The system-environment interaction is such that it leads to a full decoherence: there exists a time scale $\tau_D$, called \emph{decoherence time}, such that
asymptotically for interaction times $t\gg \tau_D$:
i) there emerges a unique, stable in time preferred basis $\ket{i}$, so called \emph{pointer basis}, in the system's Hilbert space; ii) the reduced state of the system
$\varrho_S$ becomes stable and diagonal in the preferred basis:
\be \label{decoh}
\varrho_S\equiv\tr_E\varrho_{S:E}\approx\sum_i p_i \ket{i}\bra i, 
\ee
where $p_i$'s are some probabilities and 
by $\approx$ we will always denote asymptotic equality in the deep decoherence limit $t/\tau_D\to\infty$.
We emphasize that we assume here the \emph{full decoherence}, so that the system 
decoheres in a basis rather than in higher-dimensional pointer superselection 
sectors (decoherence-free subspaces).

Coming back to the Definition \ref{obj}, we first add an important \emph{stability requirement}: 
the observers can find out the state of $S$ without perturbing it 
\emph{repeatedly and arbitrary many times}. 
In our view, this captures well  the intuitive feeling of objectivity
as something stable in time rather than fluctuating. 
Thus, if Definition \ref{obj}
is to be non-empty, it should be understood in the time-asymptotic and hence decoherence regime, 
which in turn implies that the state of $S$ which can possibly exist objectively, is determined by the decohered state
(\ref{decoh}). We will show it on a concrete example we study later.

Next, we specify the observers. Apart from the environmental ones, we also allow for a,  
possibly only hypothetical, \emph{direct observer}, who can 
measure $S$ directly. We feel such a  observer is needed
as a reference, to verify that the findings of the environmental observers are the same as if one 
had a direct access to the system.

It is clear that what the observers can determine are the eigenvalues $p_i$ 
of the decohered state (\ref{decoh})---they otherwise know the pointer basis $\ket{i}$, as if not, they would not know what the information they get is about.
Hence, the ''state'' in Definition \ref{obj}, which gains the objective existence, is the 
''classical part'' of the decohered state (\ref{decoh}), i.e. its spectrum $\{p_i\}$. 

The word ''find out'' we interpret as the observers performing von Neumann (as more informative than generalized)
measurements on their subsystems.
By the ''independence'' condition, they act independently, 
i.e. there can be no correlations between the measurements and the corresponding  
projectors must be fully product:
\ben\label{prod}
\Pi^{M_S}_i\otimes\Pi^{M_1}_{j_1}\otimes\cdots\otimes\Pi^{M_N}_{j_N},
\een
where all $\Pi$'s are mutually orthogonal Hermitian projectors, $\Pi^{M_k}_j\Pi^{M_k}_{j'}=0$ for $j\ne j'$.

Now the crucial word ''perturbation'' needs to be made precise. The debate about its meaning has been actually
at the very heart of Quantum Mechanics  from its beginnings, starting from the 
famous work of Einstein, Podolsky and Rosen (EPR) \cite{EPR} and the response of Bohr \cite{Bohr}.
It is quite intriguing that this debate appears in the context of objectivity.
The exact definitions of the EPR and Bohr notions of non-disturbance are still a subject of some debate and we adopt here
their formalizations from Ref.~\cite{Wiseman}: the sufficient condition for the EPR non-disturbance is the 
\emph{no-signaling principle}, stating that the partial state of one subsystem is insensitive
to measurements performed on the other subsystem (after forgetting the results) \cite{no-signaling}.
Quantum Mechanics obeys the no-signaling principle, but Bohr argued that the EPR's notion 
is too permissive, as it only prohibits ''mechanical'' disturbance, and proposed a stricter one, which can be 
formally stated \cite{Wiseman} that the whole \emph{joint} state 
must stay invariant under local measurements on one subsystem (after forgetting the results). 

For the purpose of this study we adopt Bohr's point of view, adapted to our particular 
situation---we assume that \emph{neither of the observers Bohr-disturbs the rest}
(in the $E\to S$ direction it is our formalization of the Definition \ref{obj}, while in the $S\to E$
it follows from the repetitivity requirement).
Together with the product structure (\ref{prod}), this implies that on each $S, E_1,\dots E_N$
there exists a \emph{non-disturbing measurement}, which leaves the whole asymptotic state  
$\varrho_{S:E}(\infty)$ of the system and the observed environment invariant (we will specify the size of the
observed environment later).
For the system $S$ it is obviously defined by the projectors on the pointer basis $\ket i$, as by assumption this is the only basis
preserved by the dynamics. For the environments we allow for a general higher-rank
projectors $\Pi^{M_k}_j$, $k=1,\dots,N$, and not necessarily spanning the whole space, 
as the environments can: i) have inner degrees of freedom not correlating
to $S$ and ii) correlate to $S$ only through some subspaces of their Hilbert spaces
(we will later encounter such a situation in the concrete example).

When more than one observer preform the non-disturbing measurements,
 a further specification of Bohr-nondisturbance is needed. Allowing for general correlations
$p_{ij_1\dots j_N}\equiv\tr[\ket i\bra i\otimes \Pi^{M_1}_{j_1}\otimes\cdots\otimes\Pi^{M_N}_{j_N}\varrho_{S:E}(\infty)]$
may lead to a disagreement: if one of the observers measures first, the ones measuring afterwards may find 
outcomes depending on the result of the first measurement (if the observers do not discard their results an meet to compare them later).
This can hardly be called objectivity and we thus add to the Definition \ref{obj} an obvious \emph{agreement requirement}:
''...observers can find out the \emph{same} state of $S$ independently,...'', leading to a natural conclusion \cite{agreement}:
\be\label{agree}
\big(\ p_{ij_1\dots j_N}\ne 0\  \text{  iff  } \ i=j_1=...=j_N\ \big) \Rightarrow p_{ii\dots i}=1,
\ee
i.e. the environmental Bohr-nondisturbing measurements must be \emph{perfectly correlated} with the pointer basis.
Hence, after forgetting the results, the asymptotic post-measurement state $\varrho^M_{S:E}(\infty)$ reads (by $\infty$ we denote $t\gg\tau_D$ asymptotic):
\ben
&& \varrho^M_{S:E}(\infty)=\sum_{i,j_1,\dots , j_N}p_{ij_1\dots j_N}\varrho_{ij_1\dots j_N}^{S:E}(\infty)=\nonumber\\
&&\sum_i \ket i\bra i\otimes{\bf \Pi}_i\,\varrho_{S:E}(\infty)\,\ket i\bra i\otimes {\bf \Pi}_i,
\een
where ${\bf \Pi}_i\equiv\Pi^{M_1}_{i}\otimes\cdots\otimes\Pi^{M_N}_{i}$.

Now we are ready for \emph{the crucial step}: we impose the relevant form of the Bohr-nondisturbance condition:
\ben\label{maxcorr}
\sum_i \ket i\bra i\otimes{\bf \Pi}_i\,\varrho_{S:E}(\infty)\,\ket i\bra i\otimes {\bf \Pi}_i=\varrho_{S:E}(\infty),
\een
whose only solution \cite{Wiseman} are the, so called, Classical-Quantum (CQ) states \cite{QC}:
\ben\label{wism}
\varrho_{S:E}(\infty)=\sum_i p_i \ket i\bra i\otimes{\bf R}^E_i,
\een
where $p_i$ are the probabilities from Eq.~(\ref{decoh}) and ${\bf R}_i^E$ are some residual states
in the space of all the environments with mutually orthogonal supports: ${\bf R}_i^E{\bf R}_{i'}^E=0$ for $i\ne i'$.
Hence, ${\bf R}^E_i$ are perfectly distinguishable \cite{NielsenChuang} through the assumed non-disturbing
measurements ${\bf \Pi}_i$, projecting on their supports.

The derived form (\ref{wism}) sheds some light on the word ''many'' in the Definition \ref{obj}: 
the compatible states (\ref{wism}) are necessarily $S:E$ separable, 
while we argue that \emph{generically}, for  large systems, the unitary system-environment evolution $U_{S:E}$ 
leads to $S:E$ entanglement (see e.g. Ref.~\cite{entanglement} for the definition of the latter). 
We first recall  that the initial states, weather pure or mixed,
are always assumed to be $S:E$ product---the system and the environment did not interact in the remote past
and there is \emph{no prior information} about the system in the environment.
The entanglement generation is then clear for pure initial states,
as entanglement is the only form of correlation for such states and without a $S:E$ correlation there 
can be no decoherence (\ref{decoh}). For mixed initial states the situation is more subtle as in finite-dimensional 
state-spaces there exist non-zero volume separable balls around the identity operator \cite{Gurvits}.
If the $S:E$ state is initially in this ball, the unitary evolution will not lead it out of it, while building enough correlations
for the decoherence (\ref{decoh}) to happen. However, for large dimensions, the radius of the largest separable ball 
decreases as $\sim 1/d$ \cite{Gurvits} and for infinite-dimensional spaces becomes strictly zero (see e.g. Ref.~\cite{Cirac}).  
This is the case here: the environment must be of a large dimension if it is to have a large informational capacity, needed
to carry a large number of copies of a state of $S$.  
Thus, the $S:E$ entanglement is generically produced during the evolution, as hitting the separable ball becomes highly unprobable due 
to its vanishing measure.
The only way then to eventually obtain a separable
state from an entangled one is by forgetting subsystems---some portions of the environment pass unobserved,
as it is actually always the case in reality. Thus, slightly abusing the language and identifying observers with the fractions of the 
environment they observe, we can interpret ''many'' as sufficiently many 
but not all---some loss of information
is necessary.  In what follows the total observed fraction of the environment will be denoted by $f$ or $fE$ (depending on the context)
and all the states above should be understood as $\varrho_{S:fE}(\infty)$.

Finally, let us look at the residual states ${\bf R}_i^E$ in Eq.~(\ref{wism}). We comeback to the demand of independent 
ability to determine the state of $S$, already used in Eq.~(\ref{prod}), and we further interpret it as a \emph{strong independence}: 
\emph{the only correlation between the environments should be the common information about the system}.
In other words, conditioned by the information about the system, there should  be  no correlations between the environments.
Thus, once one of the observers finds a particular result $i$, the conditional state should be fully product.
Since the direct observer is already uncorrelated by Eq.~(\ref{wism}), this implies that:
\be
{\bf R}_i^{fE}=\varrho^{E_1}_i\otimes\cdots\otimes \varrho^{E_{fN}}_i.
\ee
and the states $\varrho^{E_k}_i$ must be perfectly distinguishable for each environment $E_k$ 
independently:
\be\label{distng}
\varrho^{E_k}_i\varrho^{E_k}_{i'}=0\ \text{ for }\ i\ne i', 
\ee
since by the Bohr-nondisturbance (\ref{maxcorr}) for any $k$ it holds $\Pi^{M_k}_i\varrho^{E_k}_i\Pi^{M_k}_i=\varrho^{E_k}_i$  and 
$\Pi^{M_k}_i\Pi^{M_k}_{i'}=0$ for $i\ne i'$.

Gathering all the above facts together, we finally obtain:
if there is a decoherence mechanism that asymptotically leads to an objectively existing state 
of $S$ in the sense of Definition \ref{obj}, then
the asymptotic joint state of the system and the observed environment fraction 
(after the necessary tracing out of some of the environment)
must be of a special Classical-Classical \cite{oppenheim, CC} form:
\begin{equation}\label{br2}
\varrho_{S:fE}(\infty)=\sum_i p_i \ket i_S\bra i\otimes \varrho^{E_1}_i\otimes\cdots\otimes \varrho^{E_f}_i,
\end{equation}
where all $\varrho^{E_k}_i$ satisfy (\ref{distng}).

From the quantum information point of view, state (\ref{br2}) is a final state of a 
process similar to \emph{quantum state broadcasting} \cite{broadcasting}. 
The latter is a task (described by a linear map), which aims at producing from a  given state $\varrho$ a multipartite state 
$\varrho^{br}_{E_1\dots E_N}$, called an N-party broadcast state for $\varrho$, such that for every reduction
$\tr_{E_1\dots \hat E_k\dots E_N}\varrho^{br}_{E_1\dots E_N}=\varrho$, thus proliferating $\varrho$, 
but in a more subtle manner then by cloning.
Remarkably there is a weaker form of broadcasting, \emph{spectrum broadcasting} \cite{my}---a task aiming at proliferating merely a 
spectrum of a quantum state, or equivalently a classical probability distribution. We define it as follows: 
$\varrho^{s-br}_{E_1\dots E_N}$ is a spectrum broadcast state for $\varrho$, with $\text{Sp}\varrho\equiv\{p_i\}$, if for every reduction $E_k$ there exist
encoding states $\varrho^{E_k}_i$ such that:
\be
\tr_{E_1\dots \hat E_k\dots E_N}\varrho^{s-br}_{E_1\dots E_N} =\sum_ip_i\varrho^{E_k}_i \ \text{\ and }\ \varrho^{E_k}_i\varrho^{E_k}_{i'\ne i}=0
\ee 
(comparing to Ref.~\cite{my} we allow for arbitrary encoding states $\varrho^{E_k}_i$, as long as they are perfectly distinguishable).
For a given $\varrho$, a spectrum broadcast state $\varrho^{s-br}_{E_1\dots E_N}$ allows then one to locally 
recover perfect copies of the spectrum $\text{Sp}\varrho$ (through the projective measurements of the supports of $\varrho^{E_k}_i$)---the
spectrum is \emph{redundantly proliferated}. 
This is clearly the case of the state (\ref{br2}) due to the distinguishability (\ref{distng}):
(\ref{br2}) is a spectrum broadcast state for the decohered state (\ref{decoh}).
Condition (\ref{distng}) forces the correlations in (\ref{br2}) to be entirely classical
and thus the detailed structures of $\varrho^{E_k}_i$ (e.g. their ranks) become irrelevant for the correlations.
One can even pass to the purifications $\ket{\Psi^{E_k}_i}$ \cite{NielsenChuang} of  $\varrho^{E_k}_i$, which by (\ref{distng}) will be
mutually orthogonal for $i\ne i'$.
In the equivalent language of quantum channels \cite{NielsenChuang}, 
the redundant classical information transfer from the system to the observed environment
is asymptotically described by a $CC$-type channel defined by (\ref{br2}) \cite{my}.

The result (\ref{br2}) can be then re-stated as: in the presence of decoherence,
spectrum broadcasting is a necessary condition for objective existence, in the sense of Definition \ref{obj},
of the classical state of $S$ (=the spectrum of (\ref{decoh})). In other words, if a decoherence mechanism leads to a redundant production
of classical information records about the system, and hence to objectively existing classical state of $S$, 
it is necessarily achieved (in the asymptotic limit) through 
spectrum broadcasting.

Conversely, a spectrum broadcasting process resulting in a  state 
(\ref{br2}) (with the crucial property (\ref{distng})) 
leads to the objective existence in the sense of Definition \ref{obj} of the classical state $\{p_i\}$. 
Indeed, projections on the pointer basis $\ket i$ and on the disjoint supports of $\varrho^{E_k}_i$ constitute
the preferred, non-disturbing measurements. Performing them independently, 
the observers will all detect the same probability distribution $\{p_i\}$ without Bohr-disturbing the quantum state of the 
system (\ref{decoh}) and the measurements can be repeated arbitrary many times.

Summarizing, under the assumptions elaborated above, we have proven the following implications,
identifying spectrum broadcasting as the physical process responsible for the appearance of the classical objectivity:
\begin{eqnarray}
\begin{array}{c}\text{Decoherence}+\left(\begin{array}{c} \text{Objective}\\ 
\text{Existence}\end{array}\right)\Rightarrow \left(\begin{array}{c} \text{Spectrum}\\ 
\text{Broadcasting} \end{array}\right)\\
\\
\text{Objective Existence} \Leftarrow  \text{Spectrum Broadcasting } (\ref{br2})\end{array}\label{impl}
\end{eqnarray}
We also note that the form (\ref{br2}) resolves the apparent puzzle appearing  within Quantum Darwinism \cite{ZurekNature} and mentioned in the
Introduction: how can multiple information records be produced during a quantum evolution when state cloning is forbidden in quantum mechanics \cite{no_cloning}?
The answer from (\ref{br2}) is that: i) only state's spectrum  is proliferated and 
ii) instead of clones rather \emph{classically correlated copies} are produced. 

It may seem that by the time-stability requirement of objectivity, our reasoning may exclude time evolving classical objective states and
lead to the classical Zeno paradox. This is however not so. 
Moving outside the strict decoherence framework, within which our results have been derived, 
one can allow for a changing in time pointer basis $\ket{i(t)}$, and hence probabilities $p_i(t)$ (cf. Eq.~(\ref{decoh})),
but evolving on a much slower time-scale than that of the decoherence.  This is the case in most of the realistic situations,
as the decoherence time-scales are usually very short, and it opens the possibility for objectively existing, 
time-evolving classical states $\{p_i(t)\}$ iff the spectrum broadcast state (\ref{br2}) is formed fast enough for every $t$.

As a final touch, we quote the results of Refs.~\cite{ontology} on the epistemological versus ontological
interpretation of a quantum state itself: under suitable assumptions,
a state of a quantum system is a \emph{property of the system} rather than a state of knowledge 
about it. This somewhat strengthens our result and justifies the use of quantum states 
for studying objective existence: the latter gains a certain ontological status, as it intuitively should.

\section{Entropic Condition of Quantum Darwinism is not a sufficient condition for objectivity} \label{qdcond}

In the studies of Quantum Darwinism the objective existence has been so far argued based on a single
functional condition, which 
we will call \emph{Quantum Darwinism condition} (see e.g. Refs.~\cite{ZurekNature,Zurekspins,ZurekPRL} and references therein):
\be\label{Zurek}
I\left(\varrho_{S:fE}\right)=H_S,
\ee
where $I(\varrho_{AB})\equiv S_{\text{vN}}(\varrho_A)+S_{\text{vN}}(\varrho_B)-S_{\text{vN}}(\varrho_{AB})$ is the quantum mutual information, 
$S_{\text{vN}}(\varrho)\equiv -\tr(\varrho\log \varrho)$ stands for the
von Neumann entropy, and $H_S\equiv S_{\text{vN}}[\varrho_S(\infty)]=H(\{p_i\})$ is the entropy of the decohered  state (\ref{decoh}). 
Condition (\ref{Zurek}) has been shown to hold in several models, including environments comprised of photons \cite{ZurekPRL}
and spins (see e.g. Ref.~\cite{Zurekspins}).
For finite times $t$, the equality (\ref{Zurek}) is not strict and holds within some error $\delta(t)$, which defines the 
\emph{redundancy} $R_\delta(t)$ as the inverse of the smallest fraction of the environment $f_{\delta(t)}$, 
for which $I[\varrho_{S:f_{\delta(t)} E}(t)]=[1-\delta(t)]H_S$. 
When satisfied, (\ref{Zurek}) implies that the mutual information between the system
and the environment fraction is a constant function of the fraction size $f$ (up to an error $\delta$ for finite times)
and the plot of $I$ against $f$ exhibits a characteristic plateau, called the \emph{classical plateau} (see e.g. Ref.~\cite{ZurekNature}). 
The appearance of this plateau has been heuristically explained in the Quantum Darwinism literature 
as a consequence of the \emph{redundancy}: classical information about the system exists in many copies in the environment fractions and 
can be accessed independently and without perturbing the system by many observers, thus leading to objective existence of a state of $S$ \cite{ZurekNature}.
Those far reaching statements has been based only on the condition (\ref{Zurek}).

But the motivation behind using  (\ref{Zurek}) to prove the objective existence
is somewhat doubtful as it comes solely from the classical world \cite{ZurekNature}: 
in the classical information science condition (\ref{Zurek}) is equivalent to a perfect correlation of both systems \cite{ThomasCover}.
That is one system has a full information about the other and indeed in a multipartite setting this information thus exists objectively,
in accord with the Definition \ref{obj}. 
But in the quantum world the situation is very different \cite{Michal}: surprisingly, Quantum Darwinism condition (\ref{Zurek}) 
alone \emph{is not sufficient} to guarantee objectivity in the sense of Definition \ref{obj} (see also Ref.~\cite{Fields} in this context). 
It is clear that the spectrum broadcast states (\ref{br2}) 
satisfy (\ref{Zurek}), but there are also \emph{entangled} states satisfying it, 
thus violating the form (\ref{br2}), 
derived from the Definition \ref{obj} as a necessary condition for objectivity. 
As a simple example consider the following state of two qubits:
\be
\varrho_{AB}\equiv p P_{\left(a\ket{00} + b\ket{11}\right)} + (1-p)P_{\left(a\ket{01}+ b\ket{10}\right)}, 
\ee
where $P_\psi\equiv\ket\psi\bra\psi$, $p \neq 1/2$, $a=\sqrt{p}$ and $b=\sqrt{1-p}$. Then the partial state
$\varrho_B=\tilde p \ket{0}\bra{0}+(1-\tilde p)\ket{1}\bra{1}$, $\tilde p\equiv pa^2+(1-p)b^2$ is diagonal 
in the basis $\ket{0},\ket{1}$ and moreover $S_{\text{vN}}(\varrho_A)=S_{\text{vN}}(\varrho_{AB})\equiv h(p)$ 
(the binary Shannon entropy \cite{ThomasCover}), so that the Quantum Darwinism 
condition holds: $I(\varrho_{AB})=S_{\text{vN}}(\varrho_B)=H_B$, 
$H_B=h(\tilde p)$, but the systems are nevertheless entangled, which one verifies directly through
the PPT criterion \cite{PPT}.

Thus, by the results of the previous Section,  
we argue that the functional criterion (\ref{Zurek}) is not enough and the objective existence, as defined by Definition \ref{obj},
should be proven at the \emph{structural} level of quantum sates. In particular, if the spectrum broadcasting form (\ref{br2}) can be
asymptotically derived in a given model, this will guarantee the objective existence. The paradigmatic shift with respect to the earlier works on 
Decoherence Theory and Quantum Darwinism we propose here, is that the core object of the analysis should be the structure of the full  
quantum state of the system $S$ and the observed environment $fE$, 
rather than the partial state of the system only (Decoherence Theory) or information-theoretical functions (Quantum Darwinism).
Below we present such a state-based analysis and explicitly derive spectrum broadcasting states
in the emblematic example for Decoherence Theory and Quantum Darwinism: a small dielectric sphere
illuminated by photons (see e.g. Refs.~\cite{JoosZeh,GallisFleming,ZurekPRL,RiedelZurek,HornbergerSipe}).


\section{The Emblematic Example of Collisional Decoherence and Quantum Darwinism}\label{sphere}
\subsection{Basic Assumptions And Methods}

\begin{figure}[t]
\begin{center}
\includegraphics[scale=0.3]{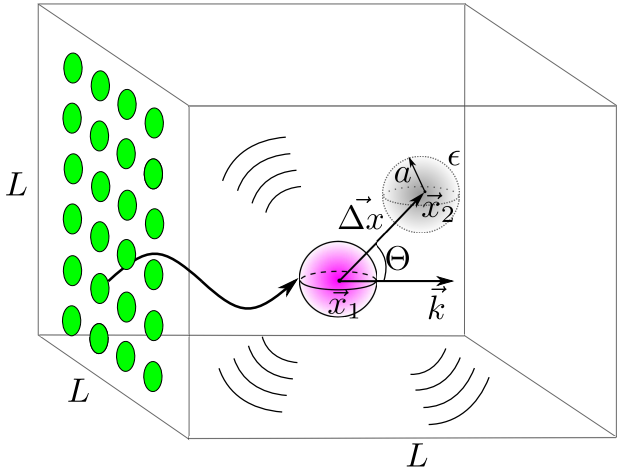}
\caption{\emph{The emblematic example of Decoherece Theory and Quantum Darwinism revisited and studied in this work.} 
A small dielectric sphere $S$ of radius $a$ and relative permittivity $\epsilon$ is illuminated by a constant flux
of photons (represented by green spots). The photons constitute the environments $E$ of the sphere.
The sphere can be at two possible locations $\vec x_1$ and $\vec x_2$,
separated by a distance $\Delta x$, much larger than the effective photon wavelengths $k\Delta x\ll 1$. 
Photons scatter elastically and slightly differently depending on where the sphere is, 
but this difference is vanishingly small for each individual scattering---the information 
about the sphere's position is diluted in the photonic environment.
However, when grouped into macroscopic fractions, the photons become collectively almost perfectly resolving
and the classical information about the sphere becomes available in the environment in multiple copies.   
We calculate the full post-scattering state of the sphere and a macroscopic fraction of the photons 
in the dipole approximation $ka\ll 1$ and show that this
redundant proliferation of information is described by spectrum broadcasting (\ref{br2}).
For technical reasons, we use box normalization: the sphere and the photons are enclosed
in a large cubic box of edge $L$ and the photon momentum eigenstates $\ket{\vec k}$ obey
periodic boundary conditions.
\label{general}}
\end{center}
\end{figure}

We first introduce the model, following the usual treatment 
(see e.g. Refs.~\cite{JoosZeh,GallisFleming,ZurekPRL,RiedelZurek}).
The system $S$ is a sphere of radius $a$ and relative permittivity $\epsilon$, 
bombarded by a constant flux of photons, which constitute the multiple environments (see Fig.~\ref{general})
and decohere the sphere.
The sphere can be located only at two positions: $\vec x_1$ or $\vec x_2$, so that
effectively its state-space is that of a qubit $\hcal_S\equiv \mathbb C^2$ with a preferred orthonormal (due to the mutual exclusiveness) 
basis $\ket{\vec x_1}$, $\ket{\vec x_2}$, which will become the pointer basis.
This greatly simplifies the analysis, yet allows the essence of the effect to be observed. 
The sphere is sufficiently massive, compared to the energy of the incoming radiation, 
so that the recoil due to the scattering photons can be totally neglected and 
photons' energy is conserved, i.e. the scattering is \emph{elastic}.
 
The environmental photons are assumed not energetic enough to individually resolve the sphere's displacement $\Delta x\equiv|\vec x_2-\vec x_1|$:
\be\label{soft}
k\Delta x\ll 1,
\ee
where $\hbar k$ is some characteristic photon momentum (the exact sens of it will be clear in what follows).
Otherwise, each individual photon would be able to resolve the position of the sphere 
and studying multiple environments would not bring anything new.
On the technical side, following the traditional approach \cite{JoosZeh,GallisFleming,ZurekPRL,RiedelZurek}, we describe the photons 
in a simplified way using box normalization: 
we assume that the sphere and the photons are enclosed in a large box of edge $L$
and volume $V=L^3$ (see Fig.~\ref{general}) and photon momentum eigenstates $\ket{\vec k}$ obey periodic boundary conditions. 
Although a more rigorous treatment was developed in Ref.~\cite{HornbergerSipe} with well localized photon states, 
we choose this traditional heuristic approach as,
at the expense of a mathematical rigor, it allows to expose the physical situation more clearly, without unnecessary mathematical details 
(we remark that the findings of  Ref.~\cite{HornbergerSipe} agree, up to an insignificant numerical factor, 
with the previous works using box normalization).
After dealing with formally divergent terms, we remove the box through the thermodynamic limit 
(signified by $\cong$) \cite{ZurekPRL,RiedelZurek}: 
\be\label{thermod}
V\to\infty, N\to\infty, \frac{N}{V}=\text{const}, 
\ee
that is we expand the box  and add more photons, keeping the photon density constant, as 
the relevant physical quantity is the radiative power, proportional
to $N/V$. The thermodynamic limit is crucial in the sense that it 
defines micro- and macroscopic regimes, which will turn to be qualitatively very distinct.

The detailed dynamics of each \emph{individual} scattering is irrelevant---the individual scatterings
are treated asymptotically in time. The interaction time $t$ enters the model differently, thought the number of scattered photons.
It may be called a ''macroscopic time''.  
Assuming photons come from the area of $L^2$ (see Fig.~\ref{general}) at a constant rate $N$ photons per volume $V$ per unit time, 
the amount of scattered photons from $t=0$ to $t$ is:
\be\label{Nt}
N_t\equiv L^2\frac{N}{V}ct,
\ee
where $c$ is the speed of light. Throughout the calculations we work with a fixed time $t$ and
pass to the asymptotic limit $t/\tau_D\to \infty$ (signified by $\approx$ or $\infty$) at the very end.

Since multiphoton scatterings can be neglected and all the photons are treated equally 
(\emph{symmetric environments}), the effective sphere-photons interaction up to time $t$ 
is of a controlled-unitary form:
\be\label{U}
U_{S:E}(t)\equiv\sum_{i=1,2}\ket{\vec x_i}\bra{\vec x_i}\otimes \underbrace{{\bf S}_i\otimes\dots\otimes {\bf S}_i}_{N_t},
\ee
where (assuming  translational invariance of the photon scattering) 
${\bf S}_i\equiv {\bf S}_{\vec x_i}=e^{-i\vec x_i\cdot \vec{\hat k}} {\bf S}_0 e^{i\vec x_i\cdot \vec{\hat k}}$ 
is the scattering matrix (see e.g. Ref.~\cite{Messiah})
when the sphere is at $\vec x_i$, ${\bf S}_0$ is the scattering matrix when the sphere is
at the origin, and $\hbar\vec{\hat k}$ is the photon momentum operator.
Due to the elastic scattering, ${\bf S}_i$'s have non-zero matrix elements only between the states $\ket{\vec k}$ 
of the same energy $\hbar c |\vec k|$.
In the sector (\ref{soft}) the interaction (\ref{U}) is vanishingly small at the level of each {\it individual} photon \cite{RiedelZurek}:  
in the thermodynamic limit ${\bf S}_1\cong{\bf S}_2$  (in a suitable sense we clarify later), 
and hence $\sum_i\ket{\vec x_i}\bra{\vec x_i}\otimes {\bf S}_i\cong {\bf 1}\otimes {\bf S}$. 
Surprisingly, this will not be true for macroscopic groups of photons.
We also note that unlike in the previous treatments \cite{JoosZeh, GallisFleming, ZurekPRL, RiedelZurek, HornbergerSipe}, 
already at this moment we explicitly include in the description \emph{all} the photons scattered up to the fixed time $t$.
Finally, the preferred role of the basis $\ket{\vec x_i}$ is already singled out now by the form of the interaction (\ref{U}) \cite{ZurekNature}.

\begin{figure}[t]
\begin{center}
\includegraphics[scale=0.35]{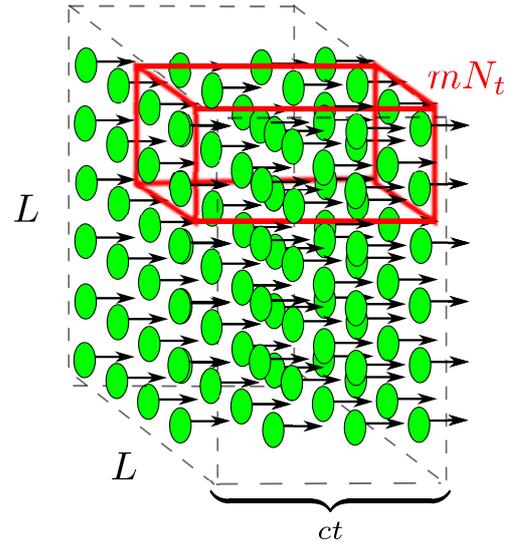}
\caption{\emph{Coarse-graining of the environment}. 
Schematic representation of a division of the whole environment---all the photons scattered in time $t$ (cf. Eq.~(\ref{Nt})), into
$M$ equal macroscopic fractions $mN_t$. Only one fraction (bounded by the red cubic cage) is shown for clarity.
The macro-fractions represent sensitivity of the detectors used to observe the scattered photons, e.g. an eye.
The exact size of the fraction given by the number $m\in [0,1]$ is irrelevant for our analysis, it is enough that it scales with the total
photon number $N_t$.
\label{division}}
\end{center}
\end{figure}

Following our critique of the Quantum Darwinism condition (\ref{Zurek}),
we analyze the model at the level of states. We need several ingredients.
First, the initial, pre-scattering ''in'' state, is as usually assumed a \emph{full product}: 
\be\label{init}
\varrho_{S:E}(0)\equiv\varrho^S_0\otimes(\varrho^{ph}_0)^{\otimes N_t},
\ee
with $\varrho^S_0$ having coherences in the preferred basis $\ket{\vec x_i}$ and $\varrho^{ph}_0$
some initial states of the photons (the environments are by assumption symmetric).
Next, we introduce a crucial \emph{environment coarse-graining} \cite{ZurekNature}:
the full environment (i.e. all the $N_t$ photons) is divided into a number of \emph{macroscopic fractions}, 
each containing $mN_t$ photons, $0\le m \le 1$ (Fig.~\ref{division}). By {\it macroscopic} 
we will always understand ''scaling with the total number of photons $N_t$''. 
By definition, these are the environment fractions accessible to the independent observers from Section~\ref{tw}. 
Such a division may seem artificial and arbitrary, as e.g. the choice of $m$ is unspecified. 
However, observe that in typical situations detectors used to monitor fractions of the environment, 
e.g. eyes, have some minimum detection thresholds---some minimum amount of radiative energy
delivered in a given time interval is needed to trigger the detection. Each macroscopic fraction $mN_t$ 
is meant to reflect that detection threshold. Its concrete value (the fraction size $m$) 
is for our analysis irrelevant---it is enough that it scales with $N_t$. 
This coarse-graining procedure is analogous to the one used e.g. in the description of liquids \cite{fluid}:
each point of a liquid (a macro-fraction $m$ here) is in reality composed of a suitable large number of 
microparticles (individual photons). It is also employed in mathematical approach to von Neumann
measurements using, so called, macroscopic observables (see e.g. Ref.~\cite{Sewell} and the references therein).
  
Thus, we divide the detailed initial state of the environment $(\varrho^{ph}_0)^{\otimes N_t}$
into $M\equiv 1/m$ macroscopic fractions:
\ben
\underbrace{\varrho^{ph}_0\otimes\dots\otimes\varrho^{ph}_0}_{N_t}&=&\underbrace{\varrho^{ph}_0\otimes\dots\otimes\varrho^{ph}_0}_{mN_t}\otimes\dots\otimes
\underbrace{\varrho^{ph}_0\otimes\dots\otimes\varrho^{ph}_0}_{mN_t}\nonumber\\
&\equiv& \underbrace{\varrho_0^{mac}\otimes\dots\otimes\varrho^{mac}_0}_{M},\label{init_mac}
\een
where $\varrho_0^{mac}\equiv (\varrho^{ph}_0)^{\otimes mN_t}$ is the initial state of each macroscopic fraction 
(\emph{macro-state} for brevity).

After all the $N_t$ photons have scattered, the asymptotic (in the sense of the scattering theory) 
''out''-state  $\varrho_{S:E}(t)\equiv U_{S:E}(t)\varrho_{S:E}(0)U_{S:E}(t)^\dagger$, is given from Eqs.~(\ref{U},\ref{init},\ref{init_mac}) by
\ben
&&\varrho_{S:E}(t)=\nonumber\\
&&\sum_{i=1,2}\langle\vec x_i |\varrho^S_0\, \vec x_i\rangle\ket{\vec x_i}\bra{\vec x_i}\otimes\underbrace{\varrho_i^{mac}(t)\otimes\dots\otimes\varrho_i^{mac}(t)}_M\label{SE1}\\
&&+\sum_{i\ne j}\langle\vec x_i |\varrho^S_0\, \vec x_j\rangle\ket{\vec x_i}\bra{\vec x_j}\otimes\underbrace{\left({\bf S}_i\varrho^{ph}_0 {\bf S}_j^\dagger\right)^{\otimes mN_t}\otimes\dots}_M\label{SE2}
\een
where 
\be\label{rho_i}
\varrho_i^{mac}(t)\equiv\left({\bf S}_i\varrho^{ph}_0 {\bf S}_i^\dagger\right)^{\otimes mN_t}, \, i=1,2. 
\ee

By the argument of Section~\ref{tw}, in order to have a chance to observe
the broadcasting state (\ref{br2}), we trace out some of the environment. 
In the current model it is important that the forgotten fraction must be \emph{macroscopic}:
we assume that $fM$, $0\leq f\leq 1$ out of all $M$ macro-fractions of Eq.~(\ref{init_mac}) are observed, 
while the rest, $(1-f)M$, is traced out. 
The resulting partial state reads 
(cf. Eqs.~(\ref{SE1},\ref{SE2})):
\ben\label{SfE}
&&\varrho_{S:fE}(t)=\sum_{i=1,2}\langle\vec x_i |\varrho^S_0\, \vec x_i\rangle\ket{\vec x_i}\bra{\vec x_i}\otimes\left[\varrho_i^{mac}(t)\right]^{\otimes fM}\label{i=j}\\
&&+\sum_{i\ne j}\langle\vec x_i |\varrho^S_0\, \vec x_j\rangle\left(\tr {\bf S}_i\varrho^{ph}_0 {\bf S}_j^\dagger\right)^{(1-f)N_t}\ket{\vec x_i}\bra{\vec x_j}\otimes\nonumber\\
&&\qquad\qquad\qquad\qquad\qquad\qquad\otimes\left({\bf S}_i\varrho^{ph}_0 {\bf S}_j^\dagger\right)^{\otimes fN_t}.
\label{i ne j}
\een

We finally demonstrate that in the soft scattering sector (\ref{soft}), 
the above state is asymptotically of the broadcast form (\ref{br2}) 
by showing that in the deep decoherence regime $t\gg \tau_D$ two effects take place:
\begin{enumerate}
\item The coherent part $\varrho_{S:fE}^{i\ne j}(t)$ given by Eq.~(\ref{i ne j}) vanishes in the trace norm:
\be\label{znikaogon}
||\varrho_{S:fE}^{i\ne j}(t)||_{\text{tr}}\equiv\tr\sqrt{\left[\varrho_{S:fE}^{i\ne j}(t)\right]^\dagger\varrho_{S:fE}^{i\ne j}(t)}\approx 0.
\ee
\item The post-scattering macroscopic states $\varrho_i^{mac}(t)$ (cf. Eq.~(\ref{rho_i})) become perfectly distinguishable:
\be\label{nonoverlap}
\varrho_1^{mac}(t)\varrho_2^{mac}(t)\approx 0,
\ee
or equivalently using the generalized overlap \cite{Fuchs}:
\ben
&&B\left[\varrho^{mac}_1(t),\varrho^{mac}_2(t)\right]\equiv\nonumber\\
&&\quad\quad\quad\equiv\tr\sqrt{\sqrt{\varrho^{mac}_1(t)}
\varrho^{mac}_2(t)\sqrt{\varrho^{mac}_1(t)}}\approx 0,\label{nonoverlap_norm}
\een
despite of the individual (microsopic) states becoming equal in the thermodynamic limit.
\end{enumerate}
The first mechanism above is the usual decoherence of $S$ by $fE$---the suppression of coherences in the preferred basis $\ket{\vec x_i}$. 
Some form of quantum correlations may still survive it, 
since the resulting state (\ref{i=j}) is generally of a Classical-Quantum (CQ) form \cite{CQ}. 
Those relict forms of quantum correlations are damped by the second mechanism: the asymptotic   
perfect distinguishability (\ref{nonoverlap}) of the post-scattering macro-states $\varrho_i^{mac}(t)$. 
Thus, the state $\varrho_{S:fE}(\infty)$ becomes of the spectrum broadcast form (\ref{br2})
for the distribution:
\ben\label{pi}
p_i=\langle\vec x_i |\varrho^S_0\, \vec x_i\rangle,
\een 
which by implications (\ref{impl}) gains objective existence in the sense of Definition \ref{obj}.

\subsection{Broadcasting Phase - Pure  Environments}
For greater transparency, we first demonstrate the mechanisms (\ref{znikaogon},\ref{nonoverlap}), and hence
a formation of the broadcast state (\ref{br2}), in a case of 
pure initial environments: 
\ben\label{pure}
\varrho_{ph}^0\equiv\ket{\vec k_0}\bra{\vec k_0},\ k_0\Delta x\ll 1, 
\een
i.e. all the photons come from the same direction
and have the same momenta $\hbar k_0$, $k_0\equiv|\vec k_0|$, satisfying (\ref{soft}). 
To show (\ref{znikaogon}), observe that $\varrho_{S:fE}^{i\ne j}(t)$, defined by Eq.~(\ref{i ne j}), is of a simple form in the basis $\ket{\vec x_i}$:
\be\label{matrix}
\varrho_{S:fE}^{i\ne j}(t)=\left[\begin{array}{cc} 0 & \gamma C\\
                                                                         \gamma^*C^\dagger & 0\end{array}\right],
\ee
where $\gamma\equiv\langle\vec x_1 |\varrho^S_0\, \vec x_2\rangle(\tr {\bf S}_1\varrho^{ph}_0 {\bf S}_2^\dagger)^{(1-f)N_t}$ and
$C\equiv ({\bf S}_1\varrho^{ph}_0 {\bf S}_2^\dagger)^{\otimes fN_t}$. Since ${\bf S}_i$'s are unitary and $\varrho^{ph}_0\geq 0$, $\tr\varrho^{ph}_0=1$,
we obtain:
\ben
&&||\varrho_{S:fE}^{i\ne j}(t)||_{\text{tr}}=\nonumber\\
&&|\gamma|\tr\left({\bf S}_1\varrho^{ph}_0 {\bf S}_1^\dagger\right)^{\otimes fN_t}+|\gamma|\tr\left({\bf S}_2\varrho^{ph}_0 {\bf S}_2^\dagger\right)^{\otimes fN_t}\\
&&=2|\langle\vec x_1 |\varrho^S_0\, \vec x_2\rangle|\left|\tr {\bf S}_1\varrho^{ph}_0 {\bf S}_2^\dagger\right|^{(1-f)N_t}\label{|i ne j|}
\een
The decoherence factor $|\tr {\bf S}_1\varrho^{ph}_0 {\bf S}_2^\dagger|^{(1-f)N_t}$ for the pure case (\ref{pure}) has been 
extensively studied before (see. e.g. Refs.~\cite{JoosZeh, GallisFleming, ZurekPRL, RiedelZurek, HornbergerSipe}). Let us briefly recall the main 
results. Under the condition (\ref{soft}) and
using the classical cross section of a dielectric sphere in the dipole approximation $k_0a\ll1$, 
one obtains in the box normalization:
\ben\label{psipsi}
&&\langle\vec k_0|{\bf S}_2^\dagger {\bf S}_1\vec k_0\rangle=1+i\frac{8\pi\Delta x k_0^5 \tilde a^6}{3L^2} \cos\Theta\nonumber\\
&&-\frac{2\pi\Delta x^2 k_0^6 \tilde a^6}{15L^2}
\left(3+11\cos^2\Theta\right)+O\left[\frac{(k_0\Delta x)^3}{L^2}\right],
\een
where $\Theta$ is the angle between the incoming direction $\vec k_0$ and the displacement vector $\vec{\Delta x}\equiv\vec x_2-\vec x_1$
and $\tilde a\equiv a [(\epsilon-1)/(\epsilon+2)]^{1/3}$. This implies:
\ben
&&\left|\tr {\bf S}_1\varrho^{ph}_0 {\bf S}_2^\dagger\right|^{(1-f)N_t}=\left|\langle\vec k_0|{\bf S}_2^\dagger {\bf S}_1\vec k_0\rangle\right|^{(1-f)N_t}\cong\nonumber\\
&&\left[1-\frac{2\pi\Delta x^2 k_0^6 \tilde a^6}{15L^2}
\left(3+11\cos^2\Theta\right)\right]^{L^2(1-f)\frac{N}{V}ct}\label{linia2}\\
&&\xrightarrow{\text{therm.}}\text e^{-\frac{(1-f)}{\tau_D}t}.\label{decay_ogon}
\een
In the second line above we used Eq.~(\ref{psipsi}) up to the leading order in $1/L$; in the last line we removed the box normalization
through the thermodynamical limit (\ref{thermod})  and thus obtained the decoherence time \cite{ZurekPRL,RiedelZurek}:
\be
{\tau_D}^{-1}\equiv\frac{2\pi}{15}\frac{N}{V}\Delta x^2 c k_0^6 \tilde a^6 \left(3+11\cos^2\Theta\right).
\ee
Eqs.~(\ref{|i ne j|},\ref{decay_ogon}) imply that $||\varrho_{S:fE}^{i\ne j}(t)||_{\text{tr}}\leq 2\text e^{-(1-f)t/\tau_D}|\langle\vec x_1 |\varrho^S_0\, \vec x_2\rangle$, 
since the sequence $(1+x/N)^N$ is monotonically increasing.
As a result, whenever we forget a macroscopic fraction of the environment ($f<1$),
the resulting coherent part $\varrho_{S:fE}^{i\ne j}(t)$ decays in the trace norm exponentially, with the characteristic
time $\tau_D/(1-f)$. This completes the first step (\ref{znikaogon}).

The asymptotic orthogonalization (\ref{nonoverlap}) is also straightforward to show in the case of pure environments.
The post-scattering states of the environment macro-fractions, Eq.~(\ref{rho_i}), are all pure:
\be
\varrho_i^{mac}(t)=\left({\bf S}_i\ket{\vec k_0}\bra{\vec k_0}{\bf S}_i^\dagger\right)^{\otimes mN_t}\equiv \ket{\Psi_i^{mac}(t)}\bra{\Psi_i^{mac}(t)},
\ee
so it is enough to consider their overlap:
\ben
&&\left|\langle\Psi_2^{mac}(t)|\Psi_1^{mac}(t)\rangle\right|=\left|\langle\vec k_0|{\bf S}_2^\dagger {\bf S}_1\vec k_0\rangle\right|^{L^2m\frac{N}{V}ct}\\
&&\xrightarrow{\text{therm.}}\text e^{-\frac{m}{\tau_D}t}.\label{ortogonalizacja}
\een
Thus, for $t\gg\tau_D$ the states of the macro-fractions $\Psi_i^{mac}(t)$ \emph{asymptotically orthogonalize} and moreover on the same timescale $\tau_D$ as
the decay of the coherent part described by Eq.~(\ref{ortogonalizacja}) (note that $0<m,f\leq 1$ so the timescales from Eqs.~(\ref{decay_ogon},\ref{ortogonalizacja})
do not differ considerably). This shows the asymptotic formation of the broadcast state (\ref{br2}) with pure encoding states $\varrho_i^{E_k}$:
\ben
&&\varrho_{S:fE}(0)=\varrho^S_0\otimes\underbrace{\varrho_0^{mac}\otimes\dots\otimes\varrho^{mac}_0}_{fM} 
\xrightarrow[\text{therm.}]{t\gg\tau_D}\varrho_{S:fE}(\infty)=\nonumber\\
&&\sum_{i=1,2}p_i\ket{\vec x_i}\bra{\vec x_i}
\otimes\underbrace{\ket{i^{mac}}\bra{i^{mac}}\otimes\dots\otimes\ket{i^{mac}}\bra{i^{mac}}}_{fM},\nonumber\\
\label{b-state}
\een
where  $p_i$ is given by Eq.~(\ref{pi}) and $\ket{i^{mac}}\equiv\ket{\Psi_i^{mac}(\infty)}$ emerges as the non-disturbing environmental basis
in the space of each macro-fraction, spanning a two-dimensional subspace, which  carries the correlation between the macro-fraction and the sphere
(this basis depends on the initial state $\ket{\vec k_0}$). Thus, the correlations become effectively among the qubits. 
The full process (\ref{b-state}) is a combination of the measurement of the system in the pointer basis
$\ket{\vec x_i}$ and spectrum broadcasting of the result, described by a CC-type channel \cite{my} :
\ben
\Lambda^{S\to fE}_\infty(\varrho_0^S)\equiv\sum_i\langle\vec x_i |\varrho^S_0\, \vec x_i\rangle\ket{i^{mac}}\bra{i^{mac}}^{\otimes fM}.
\label{L}
\een 
Quantum Darwinism condition (\ref{Zurek}) and the classical plateau  
follow now form the Eq.~(\ref{b-state}):
\be\label{QD}
I[\varrho_{S:fE}(t)]\approx H_S,
\ee
because of the conditions (\ref{znikaogon},\ref{nonoverlap_norm})
(see Appendix \ref{IHS} for the details). Thus the mutual information 
becomes asymptotically independent of the fraction $f$ (as long as it is macroscopic). 
We stress that in our analysis Eq.~(\ref{QD}) is derived as a consequence of the spectrum broadcasting.

In Quantum Darwinism simulations for finite, fixed times $t$ (see e.g. Refs.~\cite{ZurekPRL,RiedelZurek}), one can observe that
the formation of the plateau is stronger driven by increasing the time rather than the macro-fraction $f$ (keeping all other parameters equal).
This can be straightforwardly explained by looking at the Eqs.~(\ref{decay_ogon},\ref{ortogonalizacja}): the fractions $f,m$ are by definition 
at most $1$, and hence have little effect on the decay of the exponential factors, while $t$ can be arbitrarily greater than $\tau_D$, thus 
accelerating the formation of the broadcast state (\ref{b-state}).

\begin{figure}[t]
\begin{center}
\includegraphics[scale=0.32]{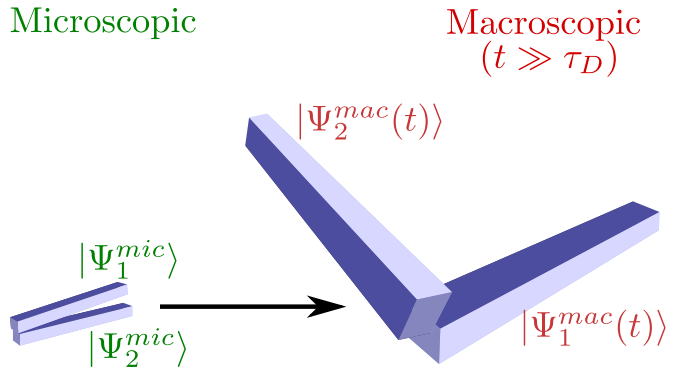}
\caption{\emph{Orthogonalization of macroscopic states}. 
At the \emph{microscopic} level, the individual post-scattering states $\ket{\Psi^{mic}_i}\equiv {\bf S}_i\ket{\vec k_0}$, 
corresponding to the sphere being at $\vec{x}_i$ 
(represented by the small solid slabs on the left)
become identical in the thermodynamic limit (cf. Eq.~(\ref{micro})) and hence completely indistinguishable.
They carry vanishingly small amount of information
about the sphere's localization, which is due to the assumed weak coupling between the sphere and each individual
environmental photon (\ref{soft}). On the other hand, the collective 
states of \emph{macroscopic} fractions $\ket{\Psi^{mac}_i(t)}\equiv \big({\bf S}_i\ket{\vec k_0}\big)^{\otimes mN_t}$ (represented by the big solid slabs on the right)
become by  Eq.~(\ref{ortogonalizacja}) more and more distinguishable in the thermodynamic (\ref{thermod}) and the deep decoherence $t\gg\tau_D$ limits. 
Together with the decoherence mechanism (\ref{znikaogon}) this leads to 
a formation of the spectrum broadcast state (\ref{br2}) with pure environmental states, 
and hence to the objective existence of the (classical) state of the sphere in the sense of Definition \ref{obj}.
\label{fig_ort}}
\end{center}
\end{figure} 

There is a very distinct difference in  the macro- and microscopic behavior of the environment, already alluded to in Refs.~\cite{ZurekPRL,RiedelZurek}.
From Eq.(\ref{psipsi}) it follows that  within the sector (\ref{soft}) the post-scattering states of {\it individual} photons (\emph{micro-states}) $\ket{\Psi^{mic}_i}\equiv {\bf S}_i\ket{\vec k_0}$, 
become identical in the thermodynamic limit and hence encode no information about the sphere's localization:
\be\label{micro}
\langle\Psi^{mic}_2|\Psi^{mic}_1\rangle\equiv \langle\vec k_0|{\bf S}_2^\dagger {\bf S}_1\vec k_0\rangle\xrightarrow{\text{therm.}}1.
\ee
This is not surprising due to the condition (\ref{soft}). On the other hand, and despite of it, 
by  Eq.~(\ref{ortogonalizacja}) macroscopic groups of photons are able to resolve the sphere's position 
and in the asymptotic limit resolve it perfectly (Fig.~\ref{fig_ort}). 
It leads to an appearance of different information-theoretical phases
in the model, which we now describe. We stress that the macro-fraction $m$
can be arbitrarily small (which only prolongs the orthogonalization time, cf. Eq.~(\ref{ortogonalizacja})), but
must scale with the total number of photons $N_t$. Indeed, for a microscopic, i.e. not scaling with $N_t$, fraction $\mu$
the limit (\ref{micro}) still holds: $[\langle\vec k_0|{\bf S}_2^\dagger {\bf S}_1\vec k_0\rangle]^\mu\xrightarrow{\text{therm.}}1$. 
Thus, if the observed portion of the environment is {\it microscopic}, the  
asymptotic post-scattering state is in fact a product one: 
\ben
&&\varrho_{S:\mu E}(0)=\varrho^S_0\otimes\left(\varrho_0^{mac}\right)^{\otimes \mu} \xrightarrow[\text{therm.}]{t\gg\tau_D}
\varrho_{S:\mu E}(\infty)=\nonumber\\
&&\sum_{i=1,2}p_i\ket{\vec x_i}\bra{\vec x_i}
\otimes\left({\bf S}_i\ket{\vec k_0}\bra{\vec k_0}{\bf S}_i^\dagger\right)^{\otimes \mu}=\\
&& \left(\sum_{i=1,2}p_i\ket{\vec x_i}\bra{\vec x_i}\right)
\otimes\ket{\Psi^{mic}}\bra{\Psi^{mic}}^{\otimes \mu},\label{product}
\een
where $\ket{\Psi^{mic}}\equiv {\bf S}_1\ket{\vec k_0} \cong {\bf S}_2\ket{\vec k_0}$ because of Eq.~(\ref{micro}) 
(and $\cong$ denotes  equality in the thermodynamic limit (\ref{thermod})). 
We call it a ''product phase'', in which $I[\varrho_{S:\mu E}(\infty)]=0$.

\begin{figure}[t]
\begin{center}
\includegraphics[scale=0.31]{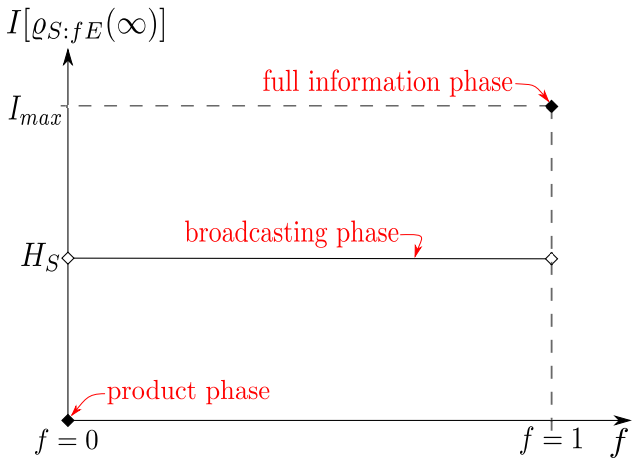}
\caption{\emph{Information-theoretical phases}. Schematic phase diagram showing three different information-theoretical phases of the model, appearing
in the thermodynamic limit (\ref{thermod}) and in the deep decoherence regime $t\gg \tau_D$. 
The horizontal axis is the macroscopic fraction $f$ of the environment $E$ under the observation.
Vertical axis represents the asymptotic mutual information between
the system $S$ and the macroscopic fraction $fE$, $I[\varrho_{S:fE}(\infty)]$.
The plot shows two phase transitions: the first one occurs at $f=0$ from the product phase of Eq.~(\ref{product})
to the broadcasting phase $0<f<1$ of Eq.~(\ref{b-state}). The second one is from the broadcasting phase
to the full information phase at $f=1$, when the observed environment is quantumly correlated with the system.
Due to the thermodynamic limit each value of the fraction $f$ should be understood modulo a microscopic fraction,
i.e. a fraction not scaling with the total photon number $N_t$ (cf. Eq.~(\ref{Nt})).
\label{phase}}
\end{center}
\end{figure}

Conversely, if we have access to the full environment, ignoring perhaps only a microscopic fraction $\mu$,
the arguments leading to Eqs.~(\ref{decay_ogon},\ref{ortogonalizacja}) do not work anymore, since from Eq.~(\ref{micro}):
\be
\left|\tr {\bf S}_1\varrho^{ph}_0 {\bf S}_2^\dagger\right|^\mu \xrightarrow{\text{therm.}} 1,
\ee
and thus there is no decoherence nor orthogonalization. The post-scattering state 
contains then the full \emph{quantum} information about the system  due to the unsuppressed system-environment entanglement produced by the controlled-unitary
interaction (\ref{U}). As a result, the mutual information attains in the thermodynamical limit
its maximum value $I_{max}$ (equal to $2H_S$ for a pure $\varrho_0^S$) and we call this regime a ''full information phase''.
We note that the rise of $I_{S:fE}$ above $H_S$ certifies the presence of entanglement \cite{entropic}.
The intermediate phase described by Eq.~(\ref{b-state}), we propose to call a ''broadcasting phase''. 
The resulting schematic phase diagram is presented in Fig.~\ref{phase}. 
The quantity experiencing discontinuous jumps is the mutual information between the system $S$ and the observed environment $fE$,
and the parameter which drives the phase transitions is the fraction size $f$.
As discussed above, each value of $f$ has to be understood modulo a micro-fraction. The appearance of the phase diagram
is a reflection of both the thermodynamic and the deep decoherence limits and its form is in agreement with the previously obtained
results (see e.g. Refs.~\cite{ZurekPRL,RiedelZurek}).

\subsection{Broadcasting Phase - Mixed Environments}
We now move to a more general case when the environmental photons are initially in a mixed state. Unlike in
the previous studies (see e.g. Refs.~\cite{JoosZeh,ZurekPRL,RiedelZurek}), we will not assume the thermal blackbody 
distribution of the photon energies, but consider a general state, diagonal in the momentum basis $\ket{\vec k}$ 
and concentrated around the energy sector (\ref{soft}):
\be\label{mu}
\varrho^{ph}_0=\sum_{\vec k} p(\vec k) \ket{\vec k}\bra{\vec k},\ \supp p\in\{\vec k:|\vec k|\Delta x\ll 1\}
\ee
As before, we work in the box normalization: the momentum eigenstates $\ket{\vec k}$ are discrete box states
and the summation is over the box modes. 
The partial post-scattering state $\varrho_{S:fE}(t)$ is given by the same Eqs.~(\ref{rho_i}-\ref{i ne j})
with the above $\varrho^{ph}_0$. The first step (\ref{znikaogon}), i.e. the decay of the coherent part,
is the same as before, as nowhere in Eqs.~(\ref{matrix}-\ref{|i ne j|}) the purity was used, but the decoherence factor
is now modified. In the leading order in $1/L$ it reads \cite{JoosZeh,ZurekPRL,RiedelZurek}:
\ben
&&\left|\tr {\bf S}_1\varrho^{ph}_0 {\bf S}_2^\dagger\right|^{(1-f)N_t}\cong\nonumber\\
&&\left[1-\frac{2\pi\Delta x^2 \tilde a^6}{15L^2}\sum_{\vec k}p(\vec k)k^6
\left(3+11\cos^2\Theta_{\vec k}\right)\right]^{(1-f)N_t}\label{bareta}\\
&&\xrightarrow{\text{therm.}}\text{exp}\left[-\frac{(1-f)}{\overline{\tau_D}}t\right],\label{decay_ogon_mix}
\een
where the modified decoherence time $\overline{\tau_D}$ is given by \cite{nota}:
\be\label{tau_D_mix}
\overline{\tau_D}^{\;-1}\equiv\frac{2\pi}{15}\frac{N}{V}\Delta x^2 c \tilde a^6 
\langle\langle k^6  \left(3+11\cos^2\Theta_{\vec k}\right)\rangle\rangle,
\ee
and $\langle\langle \cdot \rangle\rangle$ denotes the averaging with respect to $p(\vec k)$.

Completing the second step (\ref{nonoverlap_norm}) is more involved (our calculation is partially similar to that of Ref.~\cite{RiedelZurek}).
We first calculate the Bhattacharyya coefficient $B(\varrho_1,\varrho_2)$ for the individual states $\varrho^{mic}_i\equiv {\bf S}_i\varrho_0^{ph}{\bf S}_i^\dagger$. Let:
\be
\sqrt{\varrho^{mic}_1} \varrho^{mic}_2\sqrt{\varrho^{mic}_1}\equiv {\bf S}_1\left(\sum_{\vec k, \vec k''} M_{\vec k \vec k''} 
\ket{\vec k}\bra{\vec k''}\right){\bf S}_1^\dagger,\label{M0}
\ee
where:
\ben
M_{\vec k \vec k''} \equiv \sqrt{p(\vec k)p(\vec k'')}\sum_{\vec k'}p(\vec k') \langle \vec k| {\bf S}_1^\dagger {\bf S}_2 \vec k'\rangle
\langle \vec k'| {\bf S}_2^\dagger {\bf S}_1 \vec k''\rangle\nonumber.\\
\label{M}
\een
By Eq.~(\ref{mu}) it is supported in the sector (\ref{soft}), and we diagonalize it in the leading order in $1/L$. 
For that, we first decompose matrix elements $M_{\vec k \vec k''}$ in $1/L$ and keep the leading terms only. 
Let us write:
\be
{\bf S}_1^\dagger {\bf S}_2={\bf 1}-({\bf 1}-{\bf S}_1^\dagger {\bf S}_2)\equiv {\bf 1}-b.
\ee
Matrix elements of $b$ between vectors satisfying  (\ref{soft}) are of the order of $1/L$ at most. Indeed, 
by Eq.~(\ref{psipsi}) the diagonal elements 
$b_{\vec k\vec k}=1-\langle\vec k|{\bf S}_1^\dagger {\bf S}_2\vec k\rangle=O(1/L^2)$. 
The off-diagonal elements are, in turn, determined by the unitarity of ${\bf S}_1^\dagger {\bf S}_2$ and the order of the diagonal ones:
$1=|\langle\vec k| {\bf S}_1^\dagger {\bf S}_2 \vec k\rangle|^2+\sum_{\vec k'\ne \vec k}|\langle\vec k |{\bf S}_1^\dagger {\bf S}_2 \vec k'\rangle|^2
=1-O(1/L^2)+\sum_{\vec k'\ne \vec k}|b_{\vec k\vec k'}|^2$  for any fixed $\vec k$ satisfying (\ref{soft})
(there is a single sum here), where we again used Eq.~(\ref{psipsi}). Hence:
\be\label{offdiag}
\forall_{\vec k} \colon \sum_{\vec k'\ne \vec k}|b_{\vec k\vec k'}|^2=\sum_{\vec k'\ne \vec k}\left|\langle\vec k|{\bf S}_1^\dagger {\bf S}_2\vec k'\rangle\right|^2
=O\left(\frac{1}{L^2}\right). 
\ee
As a byproduct, by the above estimates in the energy sector (\ref{soft}),
${\bf S}_1\cong {\bf S}_2$ in the strong operator topology:
$||({\bf S}_1-{\bf S}_2)|\phi\rangle||^2=||b\ket{\phi}||^2\xrightarrow{\text{therm.}}0$
for any $\ket\phi$ from the subspace defined by (\ref{soft}). 
Coming back to $M_{\vec k \vec k''}$, from Eqs.~(\ref{psipsi},\ref{offdiag}) in the leading order:
\ben
M_{\vec k \vec k''}&=&p(\vec k)^2\delta_{\vec k\vec k''}-p(\vec k)^{3/2}\sqrt{p(\vec k'')}b^*_{\vec k''\vec k}\nonumber\\
&-&p(\vec k'')^{3/2}\sqrt{p(\vec k)}b_{\vec k\vec k''}+ O\left(\frac{1}{L^4}\right).\label{M'}
\een
The first term is non-negative and is of the order of unity, while the rest is of the order $1/L$ and forms a Hermitian matrix.
We can thus calculate the desired eigenvalues $m(\vec k)$ of $M_{\vec k \vec k''}$ using 
standard, stationary perturbation theory of quantum mechanics (see e.g. Ref.~\cite{Messiah}),
treating the terms with the matrix $b$ as a small perturbation.
Assuming a generic non-degenerate situation (the measure $p(\vec k)$ in Eq.~(\ref{mu}) is injective), we obtain:
\ben\label{m}
m(\vec k)=p(\vec k)^2\left(1-b^*_{\vec k\vec k}-b_{\vec k\vec k}\right)+O\left(\frac{1}{L^4}\right),
\een
and:
\ben
&&\tr\sqrt{\sqrt{\varrho^{mic}_1} \varrho^{mic}_2\sqrt{\varrho^{mic}_1}}
=\tr\sqrt M\nonumber\\
&& \cong\sum_{\vec k}p(\vec k)\sqrt{1-2\text{Re}b_{\vec k\vec k}}\cong\sum_{\vec k}p(\vec k)\left(1-\text{Re}b_{\vec k\vec k}\right)\\
&&=1+\frac{1}{2}\sum_{\vec k}\left(\frac{M_{\vec k\vec k}}{p(\vec k)}-p(\vec k)\right)=\frac{1}{2}
+\sum_{\vec k}\frac{p(\vec k)}{2}\left|\langle\vec k|{\bf S}_1^\dagger {\bf S}_2\vec k\rangle\right|^2\nonumber\\
&&+\sum_{\vec k}\sum_{\vec k'\ne \vec k}\frac{p(\vec k)}{2}\left|\langle\vec k|{\bf S}_1^\dagger {\bf S}_2\vec k'\rangle\right|^2
\equiv 1-\frac{\bar\eta-\bar\eta'}{L^2},\label{rhorho}
\een
where we have used Eqs.~(\ref{M0},\ref{m},\ref{psipsi},\ref{M}) in the respective order, and introduced:
\ben
&&\bar\eta\equiv \frac{L^2}{2}\left(1-\sum_{\vec k}p(\vec k)\left|\langle\vec k|{\bf S}_1^\dagger {\bf S}_2\vec k\rangle\right|^2\right)
\cong \left(\overline{\tau_D}c\frac{N}{V}\right)^{-1}\label{eta}\\
&&\bar\eta'\equiv \frac{L^2}{2}\sum_{\vec k}\sum_{\vec k'\ne \vec k}p(\vec k)\left|\langle\vec k|{\bf S}_1^\dagger {\bf S}_2\vec k'\rangle\right|^2
\label{eta'}
\een
(in Eq.~(\ref{eta}) we have used Eqs.~(\ref{bareta},\ref{tau_D_mix})). This implies for the micro-states:
\ben
B\left(\varrho^{mic}_1,\varrho^{mic}_2\right) = 1-\frac{\bar\eta-\bar\eta'}{L^2}\xrightarrow{\text{therm.}}1,
\label{ort_micmix}
\een
since $\bar\eta,\bar\eta'$ are of the order of unity in $1/L$ by Eqs.~(\ref{offdiag},\ref{eta}).
Thus, under (\ref{soft}), the states $\varrho^{mic}_1,\varrho^{mic}_2$ become equal. 
This is the mixed stated analog of Eq.~(\ref{micro}), employing the generalized overlap $B(\varrho_1,\varrho_2)$.

Passing to the macro-states $\varrho_i^{mac}(t)\equiv({\bf S}_i\varrho^{ph}_0{\bf S}_i^\dagger)^{\otimes mN_t}$ 
(cf. Eq.~(\ref{rho_i})), we in turn obtain:
\ben
&&B\left[\varrho^{mac}_1(t),\varrho^{mac}_2(t)\right]=
\left(\tr\sqrt{\sqrt{\varrho^{mic}_1} \varrho^{mic}_2\sqrt{\varrho^{mic}_1}}\right)^{mN_t}\nonumber\\
&&\cong \left(1-\frac{\alpha\bar\eta}{L^2}\right)^{mN_t}\xrightarrow{\text{therm.}}
\text{exp}\left[-\frac{\alpha m}{\overline{\tau_D}}t\right],\label{ort_mix}
\een
where \cite{RiedelZurek}:
\be\label{alpha}
\alpha\equiv\frac{\bar\eta-\bar\eta'}{\bar\eta}
\ee
and we have used Eq.~(\ref{eta}).
Thus, whenever $\alpha\ne 0$, the macroscopic states satisfy 
$B[\varrho^{mac}_1(t),\varrho^{mac}_2(t)]\approx 0$ for $t\gg\overline{\tau_D}/\alpha$, despite Eq.~(\ref{ort_micmix}). 
That is, they become supported on orthogonal subspaces and hence perfectly distinguishable through orthogonal projectors 
on their supports \cite{Fuchs}.
The latter are within the subspaces $\text{span}\{\ket{\vec k} :\vec k\in \supp p\}^{\otimes mN_t}$ (cf. Eq.~(\ref{mu})),
rotated by ${\bf S}_1^{\otimes mN_t}$ and ${\bf S}_2^{\otimes mN_t}$ respectively.  
This shows the asymptotic formation of the spectrum broadcasting state (\ref{br2}):
\be\label{b-state_mix}
\varrho_{S:fE}(\infty)=\sum_{i=1,2}p_i\ket{\vec x_i}\bra{\vec x_i}
\otimes\left[\varrho_i^{mac}(\infty)\right]^{\otimes fM}
\ee
with  $\varrho^{mac}_1(\infty)\varrho^{mac}_2(\infty)=0$, and hence the objective existence
in the sense of Definition \ref{obj} of the classical state (\ref{pi}) of the sphere for the mixed environments (\ref{mu}).
Thus, all our previous pure-case findings 
apply equally well here too: for $\alpha\ne 0, f\ne 0,1$ we asymptotically observe the broadcasting phase (\ref{b-state_mix}) and 
recover the Quantum Darwinism condition (\ref{Zurek}) by the same Eq.~(\ref{QD}) (see Appendix~\ref{IHS} for the details). 
Moreover, from Eqs.~(\ref{ort_micmix},\ref{ort_mix}), all the pure-case considerations regarding micro- and macro-regimes 
(cf. Eq. (\ref{micro}) and the following paragraphs) hold true and the same phase diagram of Fig.~\ref{phase} emerges. 
This is a deep feature of the model.

However, there is one remarkable difference with respect to the pure case. 
Comparing Eqs.~(\ref{decay_ogon_mix}) and (\ref{ort_mix}) one sees that in the mixed case the timescales of decoherence (\ref{znikaogon})
and distinguishability (\ref{nonoverlap}) are a priori different: $\overline{\tau_D}$ and $\overline{\tau_D}/\alpha$ respectively.
Since $0\leq\alpha\leq 1$ the latter time is in general larger and the broadcast state is fully formed for $t \gg \overline{\tau_D}/\alpha$.
Mixedness of the environment thus slows down the process of formation of the broadcast state. 
If the difference $\overline{\tau_D}/\alpha-\overline{\tau_D}$ is sufficiently large, then for the intermediate times $\overline{\tau_D}\ll t<\overline{\tau_D}/\alpha$
the state $\varrho_{S:fE}(t)$ is approximately a CQ state, whose mutual information is given by the
Holevo quantity \cite{Holevo}: $I[\varrho_{S:fE}(t)]=S_{\text{vN}}\left[\sum_i p_i \varrho_i^{mac}(t)^{\otimes fM}\right]
-(fM)\sum_i p_i S_{\text{vN}}[\varrho_i^{mac}(t)]$.

Those different time scales were already discovered and discussed in Ref.~\cite{RiedelZurek}, where $\alpha$ was called the ''environment receptivity''
and $\alpha/\overline{\tau_D}$ the ''redundancy rate''. However, the presented physical interpretations of those quantities were rather heuristic, based 
loosely on the Quantum Darwinism condition (\ref{Zurek}) and not grounded in the full state analysis, as we have presented above. 
Moreover, the measure $p(\vec k)$ studied in Ref.~\cite{RiedelZurek} was of a special, product form: $p(\vec k)=p_{th}(k)(1/\Delta\Omega)$,
where $p_{th}(k)$ is the thermal distribution of the energies and the photons were assumed to come from a portion of the ''celestial sphere'' of 
an angular measure $\Delta\Omega$.
Above, we have shown the effect for a general, diagonal in the momentum eigenbasis state (\ref{mu}).  
Let us recall after  Refs.~\cite{ZurekPRL,RiedelZurek} that for an isotropic illumination when $p(\vec k)\equiv p(k)(1/4\pi)$ (all the directions are equally probable), 
$\alpha=0$ \cite{alpha} and there is no broadcasting of the classical information: perfectly mixed directional states of the photons
cannot store any localization information of the sphere, neither on the micro- nor at the macro-level (cf. Eqs.~(\ref{ort_micmix},\ref{ort_mix})).

By Eqs.~(\ref{decay_ogon},\ref{ortogonalizacja}) and Eqs.~(\ref{decay_ogon_mix},\ref{ort_mix}), the asymptotic formation of the
spectrum broadcast states relies, among the other things, on the full product form of the initial state (\ref{init}) and the interaction (\ref{U}) in each block $i$.
However, from the same equations it is clear that one can allow for correlated/entangled fractions of photons, as long as they stay microscopic, i.e. do not scale with $N_t$.
The corresponding terms then factor out in front of the exponentials in Eqs. (\ref{decay_ogon},\ref{ortogonalizacja},\ref{decay_ogon_mix},\ref{ort_mix})
and the formation of the spectrum broadcast states is not affected.

\subsection{Perron-Frobenius Broadcasting - ''Singular Points'' of Decoherence}

We finish with a surprising application of the classical Perron-Frobenius Theorem \cite{PF}, leading to 
``singular points'' of decoherence. Let the initial state of the sphere be 
$\varrho_0^S=\sum_i\lambda_{0i}\ket{\phi_i}\bra{\phi_i}$. Then, in the spectrum broadcast states (\ref{b-state},\ref{b-state_mix})
there appears a  (unitary-)stochastic matrix $P_{ij}(\phi)\equiv|\langle\phi_i|\vec x_j\rangle|^2$ (cf. Eq.~(\ref{pi})).
By the Perron-Frobenius Theorem it possesses at least one stable probability distribution $\lambda_{*i}(\phi)$:
$\sum_j P_{ij}(\phi)\lambda_{*j}(\phi)=\lambda_{*i}(\phi)$ and such a distribution exists
for \emph{any} initial eigenbasis $\ket{\phi_i}$ of $S$. Let us now choose it as the spectrum of the initial state $\varrho_0^S$:
$\lambda_{0i}=\lambda_{*i}(\phi)$. Then, the scattering process (\ref{U}) not only leaves this distribution unchanged, but 
broadcasts it into the environment:
\ben
&&\left[\sum_i\lambda_{*i}(\phi)\ket{\phi_i}\bra{\phi_i}\right]\otimes\left(\varrho_0^{mac}\right)^{\otimes fM} 
\xrightarrow[\text{therm.}]{t\gg\tau_D}\varrho_{S:fE}(\infty)=\nonumber\\
&&=\sum_{i}\left(\sum_jP_{ij}(\phi)\lambda_{*j}(\phi)\right)\ket{\vec x_i}\bra{\vec x_i}
\otimes\left(\varrho_i^{mac}\right)^{\otimes fM}\nonumber\\
&&=\sum_{i}\lambda_{*i}(\phi)\ket{\vec x_i}\bra{\vec x_i}
\otimes\left(\varrho_i^{mac}\right)^{\otimes fM}.
\een
The initial spectrum does not ''decohere''---that is why we have called it a ''singular point'' of decoherence.
This Perron-Frobenius broadcasting process, first introduced in Ref.~\cite{my}, 
can thus be  used to faithfully (in the asymptotic limit above) broadcast the classical message $\{\lambda_{*i}(\phi)\}$ through the
environment macro-fractions.

\section{Concluding Remarks}
In this work we have identified spectrum broadcasting of Ref.~\cite{my}, a significantly weaker form of quantum state broadcasting,
as the fundamental quantum process, which leads to objectively existing classical information. More specifically, adopting the multiple environments paradigm,
the suitable definition of objectivity (Definition \ref{obj}), and Bohr's notion of non-disturbance, we have 
proven that the only possible process which makes transition from quantum state information to 
the classical objectivity is spectrum broadcasting. 
This process constitutes a formal framework and a physical foundation for the Quantum Darwinism model, 
which, as we have pointed out, in its information-theoretical form does not produce a sufficient condition for objectivity,
since it allows for entanglement.
We have shown that in the presence of decoherence, spectrum broadcasting is a necessary and sufficient condition for the objective existence 
of a classical state of the system. It filters a quantum state and then broadcasts its spectrum  
i.e. a classical probability distribution,  in multiple copies into the environment, making it accessible to the observers.
In the picture  of quantum channels, this redundant classical information transfer from the system to the environments 
is described by a CC-type channel. 

We have illustrated spectrum broadcasting process on the emblematic example for Decoherence Theory:  
a small dielectric sphere embedded in a photonic environment. In particular, we have explicitly shown the 
asymptotic formation of a spectrum broadcasting state for both pure and general (not necessarily thermal) mixed 
photon environments. Then, we have derived in the asymptotic limit of deep decoherence 
the information-theoretical phase diagram of the model. Depending on the observed macroscopic fraction 
$f$ of the environment, it shows three phases: the product, broadcasting 
and full information phase, and is a complete agreement (up to some error $\delta$ for finite times) with the classical 
plateau of the original Quantum Darwinism studies. 
There are two phase transitions taking place: i) from the product phase to the broadcasting 
phase (at $f=0)$; ii) from the broadcasting phase ($0<f<1$) to the full information phase (at $f=1$), when the observed 
environment becomes quantumly correlated with the system. 
In addition,  we have pointed out that a special form spectrum broadcasting---the Perron-Frobenius broadcasting,
can be used to  faithfully (in the asymptotic limit) broadcast certain classical message 
through the noisy environment fractions. 

From an experimental point of view, our work opens a possibility to develop an experimentally friendly framework 
for testing Quantum Darwinism. Our central object, the broadcast state (\ref{br2}),
is in principle directly observable through e.g. quantum state tomography---a well developed, 
successful, and widely used technique. 
In contrast, the original Quantum Darwinism condition (\ref{Zurek}) relies on the quantum mutual
information and it is not clear how to measure it.

We finish with a series of general remarks and questions.

First, there is a straightforward generalization of the illuminated sphere model to a situation
where classical correlations are spectrum broadcasted \cite{my}. Consider several spheres, each with its own 
photonic environment, and separated by distances $D$ much larger than the photon wavelengths, 
$kD\gg1$ (cf. Eq.~(\ref{soft})).
The effective interaction is then a product of the unitaries (\ref{U}), e.g.:
\be
U_{S_1S_2:E_1E_2}(t)\equiv\sum_{i,j=1,2}\ket{\vec x_i}\bra{\vec x_i}\otimes\ket{\vec y_j}\bra{\vec y_j}
\otimes{\bf S}_i^{\otimes N_t}\otimes\tilde{\bf S}_j^{\otimes N_t},
\ee
for two spheres, where $\vec x_i,\vec y_j$ are the spheres' positions and ${\bf S}_i,\tilde{\bf S}_j$ are the corresponding scattering matrices,
and the asymptotic spectrum broadcast state carries now the joint probability, e.g. 
$p_{ij}\equiv\langle \vec x_i,\vec y_j|\varrho_0^S\vec x_i,\vec y_j\rangle$ (cf. Eq~(\ref{pi})). 
It is measurable by observers, who have an access to photon macro-fractions, originating from all the spheres.

Second, in the example we have studied, and in the majority of decoherence models \cite{modern}, 
the system-environment interaction Hamiltonian is of a product form:
\be\label{Hint}
H_{int}=g A_S \sum_{k=1}^N X_{E_k},
\ee
where $g$ is a coupling constant and $A_S,X_{E_1},\dots,X_{E_N}$ are some observables on the system and the environments respectively. 
The pointer basis appears then trivially as the eigenbasis of $A=\sum_i a_i \ket i\bra i$---it is arguably put by hand by the choice
of $A$.  
It is then an interesting question if there are more general interaction Hamiltonians, without a priori chosen pointer basis,
which nevertheless lead to an asymptotic formation of a spectrum broadcast state:
\be\label{br3}
\varrho_{S:fE}(t)\approx \sum_i p_i \ket{i}\bra{i}\otimes_k \varrho_i^{E_k},\  \varrho^{E_k}_i\varrho^{E_k}_{i'\ne i}=0.
\ee
Are there \emph{truly dynamical} 
mechanisms leading to stable pointer bases and objective classical states?

Viewing Eq.~(\ref{br3}) form a different angle, we note that spectrum broadcasting defines a split of information contained in the quantum state 
$\varrho_S=\sum_i p_i \ket{i}\bra{i}$ into classical and quantum parts.
As it is well known, every quantum state can be convexly decomposed in many ways into mixtures of pure states, 
so a priori such a split does not exist. Some additional process is needed. Spectrum broadcasting is an example of it: by correlating
to the preferred basis $\ket i$, it endows the corresponding probabilities $p_i=\langle i|\varrho_S| i\rangle$ with objective existence,
in the sense of Definition \ref{obj}, and defines them as a "classical part" of $\varrho_S$, leaving the states $\ket{i}\bra{i}$ as
a "quantum part" (cf. no-local-broadcasting theorem of Ref.~\cite{CC}).

Third, there appears to be a deep connection between the non-signaling principle and objective existence
in the sense of Definition \ref{obj}: the core fact that it is at all possible for 
observers to determine \emph{independently} the classical state
of the system is guaranteed by the non-signaling principle: $\tr({\bf 1}\otimes \Pi_E\varrho)=\tr_E(\Pi_E\varrho_E)$.
There is no contradiction with the Bohr-nondisturbance, as the latter is a strictly \emph{stronger} condition
than the non-signaling \cite{Wiseman}(this is the core of Bohr's reply \cite{Bohr} to EPR ). 
In fact, the above  connection reaches deeper than quantum mechanics. In a general theory, where it is
possible to speak of probabilities $p(ij|MN)$ of obtaining results $i,j$ when performing measurements $M,N$ (however defined), 
whatever the definition of objective existence may be, the requirement of the
\emph{independent} ability to locally determine probabilities by each party seem indispensable. 
This is guaranteed in the non-signaling theories, where all $p(ij|MN)$'s have well defined marginals.
In this sense non-signaling seems a \emph{prerequisite of cognition}. 
This connection will be the subject of a further research.

Finally, one may speculate on a relevance of our results for life processes. 
Already in 1961, Wigner tried to argue that the standard quantum formalism 
does not allow for the self-replication of biological systems \cite{Wigner}. 
It seemed to be confirmed by the famous no cloning theorem \cite{no_cloning}. However, now we see that 
cloning is not the only possibility. As we have shown, spectrum broadcasting 
implies a redundant replication of classical information in the environment. 
This is indispensable for the existence of life: 
one of the most fundamental processes of life is
Watson-Crick alkali encoding of genetic information into the DNA molecule and self-replication of the DNA information. 
It cannot be thus a priori excluded that spectrum broadcasting may 
indeed open a ''classical window'' for life processes within quantum mechanics.

\acknowledgements
This research is supported by ERC Advanced Grant QOLAPS and  
National Science Centre project Maestro DEC-2011/02/A/ST2/00305. We thank M. Piani for discussions
on strong independence. P.H. and R.H. acknowledge discussions with K. Horodecki, M. Horodecki, and K. \.Zyczkowski. 

\appendix
\section{Derivation of the quantum darwinism  relation (\ref{QD})}
\label{IHS}
Here we present an independent derivation of the Quantum Darwinism condition (\ref{Zurek})
for the illuminated sphere model from Section~\ref{sphere} (cf. Eq.~(\ref{QD})). 
Although illustrated on a concrete model, our derivation is indeed more general: instead of
a direct, asymptotic calculation of the mutual information $I[\varrho_{S:fE}(t)]$ in the model 
(cf. Refs.~\cite{ZurekPRL,RiedelZurek}), we will show that Eq.~(\ref{Zurek}) follows from the
mechanisms of i) decoherence, Eq.~(\ref{znikaogon}), and ii) distinguishability, Eq.~(\ref{nonoverlap_norm}),  
once they are proven.

Let the post-interaction $S:fE$ state for a fixed, finite box $L$ and time $t$ be $\varrho_{S:fE}(L,t)$.
It is given by Eqs.~(\ref{i=j},\ref{i ne j}) and now we 
explicitly indicate the dependence on $L$ in the notation. Then:
\ben
&&\left|H_S-I\left[\varrho_{S:fE}(L,t)\right]\right|\leq\nonumber\\
&&\left|I\left[\varrho_{S:fE}(L,t)\right]-I\left[\varrho^{i=j}_{S:fE}(L,t)\right]\right|\label{coh}\\
&&+\left|H_S-I\left[\varrho^{i=j}_{S:fE}(L,t)\right]\right|,\label{ort}
\een
where $\varrho^{i=j}_{S:fE}(L,t)$ is the decohered part of $\varrho_{S:fE}(L,t)$, given by Eq.~(\ref{i=j}).
We first bound the difference (\ref{coh}), decomposing the mutual information using conditional information 
$S_{\text{vN}}(\varrho_{S:fE}|\varrho_{fE})\equiv S_{\text{vN}}(\varrho_{S:fE})-S_{\text{vN}}(\varrho_{fE})$:
\be
I(\varrho_{S:fE})=S_{\text{vN}}\left(\varrho_S\right)-S_{\text{vN}}\left(\varrho_{S:fE}|\varrho_{fE}\right),
\ee
so that:
\ben
&&\left|I\left[\varrho_{S:fE}(L,t)\right]-I\left[\varrho^{i=j}_{S:fE}(L,t)\right]\right|\leq\nonumber\\
&&\left|S_{\text{vN}}\left[\varrho_{S}(L,t)\right]-S_{\text{vN}}\left[\varrho^{i=j}_{S}(L,t)\right]\right|+\label{SS}\\ 
&&\Big|S_{\text{vN}}\left[\varrho_{S:fE}(L,t)\big|\varrho_{fE}(L,t)\right]\nonumber\\
&&\quad\quad\quad\quad\quad -S_{\text{vN}}\left[\varrho_{S:fE}^{i=j}(L,t)\Big|\varrho_{fE}^{i=j}(L,t)\right]\Big|.
\label{Scond}
\een
From Eq.~(\ref{soft}), the 
total $S:fE$ Hilbert space is finite-dimensional for a finite $L,t$: there are $fN_t=$$fL^2(N/V)ct$ photons (cf. Eq.~(\ref{Nt}))
and the number of modes of each photon is approximately $(4\pi/3)(L/2\pi\Delta x)^3$. 
Hence, the total dimension is $2\times L^2f(N/V)ct\times(1/6\pi^2)(L/\Delta x)^3<\infty$ and we can use the
Fannes-Audenaert \cite{FAd} and the Alicki-Fannes \cite{FannesAlicki} inequalities to bound
(\ref{SS}) and (\ref{Scond}) respectively. For (\ref{SS}) we obtain:
\ben
&&\left|S_{\text{vN}}\left[\varrho_{S}(L,t)\right]-S_{\text{vN}}\left[\varrho^{i=j}_{S}(L,t)\right]\right|\nonumber\\
&&\leq\frac{1}{2}\epsilon_E(L,t)\log(d_S-1)+h\left[\frac{\epsilon_E(L,t)}{2}\right],
\een
where $h(\epsilon)\equiv-\epsilon\log\epsilon-(1-\epsilon)\log(1-\epsilon)$ is the binary Shannon entropy
and:
\ben
&&\epsilon_E(L,t)\equiv ||\varrho_{S}(L,t)-\varrho^{i=j}_{S}(L,t)||_{tr}\\
&&=||\varrho^{i\ne j}_{S}(L,t)||_{tr}\cong 2|c_{12}|\left[1-\frac{1}{c\overline{\tau_D}L^2}\left(\frac{N}{V}\right)^{-1}\right]^{L^2\frac{N}{V}ct}
\label{ELt}
\een
with $c_{12}\equiv\langle\vec x_1 |\varrho^S_0\, \vec x_2\rangle$, 
where we have used the reasoning (\ref{matrix}-\ref{decay_ogon}), or (\ref{decay_ogon_mix}-\ref{tau_D_mix}) for the mixed environments, 
but with $f=0$. For (\ref{Scond}) the same reasoning and the Alicki-Fannes inequality give:
\ben
&&\left|S_{\text{vN}}\left[\varrho_{S:fE}(L,t)\big|\varrho_{fE}(L,t)\right]
-S_{\text{vN}}\left[\varrho_{S:fE}^{i=j}(L,t)\big|\varrho_{fE}^{i=j}(L,t)\right]\right|\nonumber\\
&&\leq4\epsilon_{fE}(L,t)\log d_S+2 h\left[\epsilon_{fE}(L,t)\right],
\een
with:
\ben
\epsilon_{fE}(L,t)&\equiv& ||\varrho_{S:fE}(L,t)-\varrho^{i=j}_{S:fE}(L,t)||_{tr}\\
&=&||\varrho^{i\ne j}_{S:fE}(L,t)||_{tr}\\
&\cong& 2|c_{12}|\left[1-\frac{1}{c\overline{\tau_D}L^2}\left(\frac{N}{V}\right)^{-1}\right]^{L^2(1-f)\frac{N}{V}ct}.\label{EfLt}
\een
Above $L,t$ are big enough so that $\epsilon_E(L,t), \epsilon_{fE}(L,t)<1$. 
Eqs.~(\ref{SS}-\ref{EfLt}) give an upper bound on the difference (\ref{coh}) in terms of the 
decoherence speed (\ref{znikaogon}).

To bound the "orthogonalization" part (\ref{ort}), we note that since $\varrho^{i=j}_{S:fE}(L,t)$ is a CQ-state
(cf. Eq.~(\ref{i=j})), its mutual information is given by the Holevo quantity \cite{Holevo}:
\ben
I\left[\varrho^{i=j}_{S:fE}(L,t)\right]=\chi\left\{p_i,\varrho_i^{mac}(t)^{\otimes fM}\right\},
\een
where $p_i$ is given by Eq.~(\ref{pi}).
From the Holevo Theorem it is bounded by \cite{Holevo}:
\be\label{hol}
I_{max}(t)\leq\chi\left\{p_i,\varrho_i^{mac}(t)^{\otimes fM}\right\}\leq H\left(\{p_i\}\right)\equiv H_S,
\ee
where $I_{max}(t)\equiv \max_{\mathcal E}I[p_i\pi^{\mathcal E}_{j|i}(t)]$ is the fixed time maximal mutual information, extractable 
through generalized measurements $\{\mathcal E_j\}$ on the  ensemble 
$\{p_i,\varrho_i^{mac}(t)^{\otimes fM}\}$, and the conditional probabilities read:
\be\label{piE}
\pi^{\mathcal E}_{j|i}(t)\equiv\tr[\mathcal E_j\varrho_i^{mac}(t)^{\otimes fM}]
\ee
(here and below $i$ labels the states, while $j$ the measurement outcomes).
We now relate $I_{max}(t)$ to the generalized overlap   
$B\left[\varrho_1^{mac}(t)^{\otimes fM},\varrho_2^{mac}(t)^{\otimes fM}\right]$ (cf. Eq.~(\ref{nonoverlap_norm})),
which we have calculated in Eq.~(\ref{ort_mix}). Using the method of Ref.~\cite{Fuchs},
slightly modified to unequal a priori probabilities $p_i$, we obtain for an arbitarry measurement $\mathcal E$:
\ben
&&I\left(\pi^{\mathcal E}_{j|i}p_i\right)=I\left(\pi^{\mathcal E}_{i|j}\pi^{\mathcal E}_j\right)
=H\left(\{p_i\}\right)-\sum_{j=1,2}\pi^{\mathcal E}_jh\left(\pi^{\mathcal E}_{1|j}\right)\nonumber\\
&&\\
&&\geq H\left(\{p_i\}\right)-2\sum_{j=1,2}\pi^{\mathcal E}_j \sqrt{\pi^{\mathcal E}_{1|j}\left(1-\pi^{\mathcal E}_{1|j}\right)}\\
&&=H\left(\{p_i\}\right)-2\sqrt{p_1p_2}\sum_{j=1,2}\sqrt{\pi^{\mathcal E}_{j|1}\pi^{\mathcal E}_{j|2}},
\een
where we have first used Bayes Theorem $\pi^{\mathcal E}_{i|j}=(p_i/\pi^{\mathcal E}_j)\pi^{\mathcal E}_{j|i}$,
$\pi^{\mathcal E}_j\equiv\sum_i\pi^{\mathcal E}_{j|i}p_i=\tr(\mathcal E_j\sum_i\varrho_i)$, then the fact that we have only two states:
$\pi^{\mathcal E}_{2|j}=1-\pi^{\mathcal E}_{1|j}$, so that $H(\pi^{\mathcal E}_{\cdot|j})=h(\pi^{\mathcal E}_{1|j})$,
and finally $h(p)\leq2\sqrt{p(1-p)}$. On the other hand, 
$B(\varrho_1,\varrho_2)=\min_{\mathcal E}\sum_j\sqrt{\pi^{\mathcal E}_{j|1}\pi^{\mathcal E}_{j|2}}$ \cite{Fuchs}. Denoting the optimal
measurement by $\mathcal E_*^B(t)$ and recognizing that $H(\{p_i\})=H_S$, we obtain:
\ben
&&I_{max}(t)\geq I\left[p_i\pi^{\mathcal E_*^B(t)}_{j|i}(t)\right]\geq H_S-\\
&&-2\sqrt{p_1p_2}\,B\left[\varrho_1^{mac}(t)^{\otimes fM},\varrho_2^{mac}(t)^{\otimes fM}\right]\\
&&=H_S-2\sqrt{p_1p_2}\,B\left[\varrho^{mac}_1(t),\varrho^{mac}_2(t)\right]^{fM} 
\een
Inserting the above into the bounds (\ref{hol}) gives the desired upper bound on the difference (\ref{ort}):
\ben
&&\left|H_S-I\left[\varrho^{i=j}_{S:fE}(L,t)\right]\right|\leq
2\sqrt{p_1p_2}\,B\left[\varrho^{mac}_1(t),\varrho^{mac}_2(t)\right]^{fM}\nonumber\\
&&
\een
where the generalized overlap is given by Eq.~(\ref{ort_mix}): 
\ben
&&B\left[\varrho^{mac}_1(t),\varrho^{mac}_2(t)\right]\cong\nonumber\\
&&\quad\quad\quad\left[1-\frac{\alpha}{c\overline{\tau_D}L^2}\left(\frac{N}{V}\right)^{-1}\right]^{L^2m\frac{N}{V}ct}.
\label{Bh}
\een

Gathering all the above facts together finally leads to a bound on   
$\left|H_S-I\left[\varrho_{S:fE}(L,t)\right]\right|$ in terms of the speed of 
i) decoherence (\ref{znikaogon}) and ii) distinguishability (\ref{nonoverlap_norm}):
\ben
&&\left|H_S-I\left[\varrho_{S:fE}(L,t)\right]\right|\leq h\left[\frac{\epsilon_E(L,t)}{2}\right]+
 2h\left[\epsilon_{fE}(L,t)\right]+\nonumber\\
&&\label{gen1}\\
&&4\epsilon_{fE}(L,t)\log 2+2\sqrt{p_1p_2}\,B\left[\varrho^{mac}_1(t),\varrho^{mac}_2(t)\right]^{fM},\label{gen2}
\een
where $\epsilon_E(L,t)$, $\epsilon_{fE}(L,t)$, $B\left[\varrho^{mac}_1(t),\varrho^{mac}_2(t)\right]$
are given by Eqs.~(\ref{ELt}), (\ref{EfLt}), and (\ref{Bh}) respectively. Choosing
$L,t$ big enough so that $\epsilon_E(L,t),\epsilon_{fE}(L,t)\leq 1/2$ (when the binary 
entropy $h(\cdot)$ is monotonically increasing), we remove the unphysical box and obtain an estimate on
the speed of convergence of $I\left[\varrho_{S:fE}(L,t)\right]$ to $H_S$:
\ben
&&\lim_{L\to\infty}\left|H_S-I\left[\varrho_{S:fE}(L,t)\right]\right|\leq 
h\left(|c_{12}|\text e^{-\frac{t}{\overline{\tau_D}}}\right)\\
&&+2h\left(2|c_{12}|\text e^{-\frac{(1-f)}{\overline{\tau_D}}t}\right)
+8|c_{12}|\text e^{-\frac{(1-f)}{\overline{\tau_D}}t}\log 2\\
&&+2\sqrt{p_1p_2}\text e^{-\frac{\alpha f}{\overline{\tau_D}}t}.
\een
This finishes the derivation of the Quantum Darwinism condition (\ref{QD}).

We note that the result (\ref{gen1},\ref{gen2}) is in fact a general statement, valid 
in any model where: i) the system $S$ is effectively a qubit; ii) the system-environment
interaction is of a environment-symmetric controlled-unitary type:
\begin{theorem}
Let a two-dimensional quantum system $S$ interact with $N$ identical environments, each described by
a finite-dimensional Hilbert space, through a controlled-unitary interaction:
\be
U(t)\equiv\sum_{i=1,2}\ket i\bra i\otimes U_i(t)^{\otimes N}.
\ee
Let the initial state be $\varrho_{S:E}(0)=\varrho_0^S\otimes(\varrho_0^E)^{\otimes N}$ 
and $\varrho_{S:E}(t)\equiv U(t)\varrho_{S:E}(0)U(t)^\dagger$. Then for any $0<f<1$ and $t$ big enough:
\ben
&&\left|H(\{p_i\})-I\left[\varrho_{S:fE}(t)\right]\right|\leq h\left[\frac{\epsilon_E(t)}{2}\right]+
 2h\left[\epsilon_{fE}(t)\right]+\nonumber\\
&&\label{gen1}\\
&&4\epsilon_{fE}(t)\log 2+2\sqrt{p_1p_2}\,B\left[\varrho_1(t),\varrho_2(t)\right]^{fN},\label{gen3}
\een
where: 
\ben
&&p_i\equiv\langle i|\varrho_0^S|i\rangle,\,\varrho_i(t)\equiv U_i(t)\varrho_0^EU_i(t)^\dagger,\\
&&\epsilon_E(t)\equiv ||\varrho_{S}(t)-\varrho^{i=j}_{S}||_{tr},\\
&&\epsilon_{fE}(t)\equiv ||\varrho_{S:fE}(t)-\varrho^{i=j}_{S:fE}(t)||_{tr}.
\een
\end{theorem}


\begin{thebibliography}{99}
\bibitem{alpha} To prove it, we calculate $\sum_{\vec k,\vec k'}p(\vec k)\left|\langle\vec k|{\bf S}_1^\dagger {\bf S}_2\vec k'\rangle\right|^2$
for $p(\vec k)\equiv p(k)(1/4\pi)$. Or more precisely, since we are working in the box normalization, the measure  
is $p(k)(1/\Omega_k)$, where $\Omega_k$ is the number of the discrete box states $\ket{\vec k}$ 
with the fixed length $k=|\vec k|$. In the continuous limit $\Omega_k$ approaches $4\pi k^2$.
As the scattering is by assumption elastic, matrix elements $\langle\vec k|{\bf S}_1^\dagger {\bf S}_2\vec k'\rangle$ are non-zero
only for the equal lengths $k=k'$ and hence:
\be\label{sectors}
{\bf S}_1^\dagger {\bf S}_2\equiv\bigoplus_k U_k,\  U_k^\dagger U_k={\bf 1}_k.
\ee
Decomposing the summations over  $\vec k,\vec k'$ into the sums over the lengths $k,k'$ and the directions $\vec n(k), \vec n(k')$
and using (\ref{sectors}), we obtain:
\ben
&&\sum_{\vec k,\vec k'}p(\vec k)\left|\langle\vec k|{\bf S}_1^\dagger {\bf S}_2\vec k'\rangle\right|^2=
\sum_k \frac{p(k)}{\Omega_k}\sum_{\vec n(k),\vec n(k')}\left|\langle\vec k|{\bf S}_1^\dagger {\bf S}_2\vec k'\rangle\right|^2=\nonumber\\
&&\sum_k \frac{p(k)}{\Omega_k}\tr\left(P_k {\bf S}_1^\dagger {\bf S}_2 P_k {\bf S}_2{\bf S}_1^\dagger\right)
=\sum_k p(k)\frac{\tr P_k}{\Omega_k}=1,\label{1}
\een 
where $P_k\equiv\sum_{\vec n(k)}\ket{\vec k}\bra{\vec k}$ is a projector onto the subs-space of a fixed length $k$, and hence $\tr P_k=\Omega_k$. 
Comparing with Eq.~(\ref{rhorho}),  Eq.~(\ref{1}) leads to $\bar\eta-\bar\eta'=0$, and hence by definition (\ref{alpha}) to $\alpha=0$.
\end{thebibliography}
\end{document}